\documentclass[12pt]{iopart}

\usepackage{slashed}
\usepackage{bbm}
\usepackage{dsfont}
\usepackage{amsfonts}
\usepackage{bm}
\usepackage{amssymb}
\usepackage{bbold}

\begin{document}

\title[]{A BLACK HOLE SOLUTION OF HIGHER-DIMENSIONAL WEYL$-$YANG$-$KALUZA$-$KLEIN THEORY}

\author{Halil Kuyrukcu}

\address{Physics Department, Zonguldak B\"{u}lent Ecevit University, Zonguldak, 67100,
Turkey}
\ead{kuyrukcu@beun.edu.tr}
\vspace{10pt}
\begin{indented}
\item[]
\end{indented}

\begin{abstract}
We consider the Weyl$-$Yang gauge theory of gravitation in a $(4+3)$-dimensional curved space-time within the scenario of the non-Abelian Kaluza$-$Klein theory for the source and torsion-free limits. The explicit forms of the field equations containing a new spin current term and the energy-momentum tensors in the usual four dimensions are obtained through the well-known dimensional reduction procedure. In this limit, these field equations admit (anti-)dyon and  magnetic (anti-)monopole solutions as well as non-Einsteinian solutions in the presence of a generalized Wu$-$Yang ansatz and some specific warping functions when the extra dimensions associated with the round and squashed three-sphere $S^{3}$ are, respectively, included. The (anti-)dyonic solution has similar properties to those of the Reissner$-$Nordstr\"{o}m$-$de Sitter black holes of the Einstein$-$Yang$-$Mills system. However, the cosmological constant naturally appears in this approach, and it associates with the constant warping function as well as the three-sphere radius. It is demonstrated  that not only the squashing parameter $\ell$  behaves as the constant charge but also its sign can  determine whether the solution is a dyon/monopole or an antidyon/antimonopole. It is also shown by using the power series method that the existence of nonconstant warping function is essential for finding new exact Schwarzschild-like solutions in the considered model.
\end{abstract}


%\maketitle

\section{Introduction}
The Weyl$-$Yang (WY) alternative gravitational theory with field equations that contain second derivatives of the affine connection and third derivatives of the metric tensor obtained via the standard Palatini formalism \cite{palatini,misner,Tsamparlis,ferraris}  with vanishing torsion was first considered by Fairchild in the remarkable studies~\cite{fairchild76,fairchild77}. The work was inspired by early works conducted by Weyl~\cite{weyl}, Lanczos~\cite{lanczos}, Utiyama~\cite{utiyama},  Lichnerowicz~\cite{lichnerowicz1}, Stephenson~\cite{stephenson}, Higgs~\cite{higgs}, Kilmister and Newman~\cite{kilmister}, Loos~\cite{Loos}, Loos and Treat~\cite{Loos1}, and Yang~\cite{yang}. The theory was investigated and developed further by several different authors such as  Ramaswamy and  Yasskin~\cite{Ramaswamy}, Tseytlin~\cite{tseytlin}, Szczyrbac~\cite{Szczyrba}, Hehl \textit{et al.}~\cite{hehl89}, Maluf~\cite{Maluf91}, Rose~\cite{Rose}, and Guilfoyle and  Nolan~\cite{Guilfoyle} and  more recently  by Vassiliev~\cite{Vassiliev}, Mielke and Maggiolo~\cite{Maggiolo}, Shen~\cite{Shen09}, and Pasic and  Barakovic~\cite{Pasic}. It was also extended$/$generalized to the Abelian$/$non-Abelian Kaluza$-$Klein (KK) theories of unified gravity by Ba\c{s}kal and Kuyrukcu~\cite{halil13}$/$Kuyrukcu \cite{halil14}, and  its ability to solve various cosmological problems was  discussed  by Gerard~\cite{gerard}, Cook~\cite{cook},  Gonzalez~\cite{gonzalez}, Chen~\cite{chen}, Chen \textit{et al.}~\cite{chen1}, and Yang and Yeung~\cite{yeung}. This Yang$-$Mills (YM)-type theory actually may also be thought of as a  particular case of the Poincar\'{e} Gauge theory~\cite{daum,Blagojevic13} ``that can be considered as the standard theory of gravity with torsion'' according to  Hehl \textit{et al.}~\cite{hehl13} (for more historical notes, see, e.g., \cite{Dean}).

It is not surprising that the WY theory overcomes certain problems with the linearized field equations~\cite{aragone}, but it also suffers from various difficulties such a nonacceptable Newtonian limit for weak field approximation~\cite{fairchild76}, the nonexistence of a Birkhoff theorem~\cite{Blagojevic13},  uncertainty concerning the source of matter~\cite{fairchild76,fairchild77,Camenzind}, and a negative residue in the graviton propagator~\cite{leeneeman}. Nonetheless, this alternative gravitational theory still deserves further investigation and exploration  as  it  has  some interesting features as stated below, although the Einstein field equations are sufficient for the large scale of gravitational physics.

The most important and motivating feature of the theory lies in the fact that the third-order Einsteinian equations  possess mathematically  a much richer structure than their second-order counterparts within the well-established gravity theory of Einstein~\cite{yang}. The master equation of the theory can be written as follows:
\begin{eqnarray}\label{alter}
 D{\star{ \mathcal{R}_{\mu\nu}}}\equiv E_{\mu}\wedge{\star{D}}\mathcal{R}_{\nu}
- E_{\nu}\wedge{\star{ D}} \mathcal{R}_{\mu}=0.
\end{eqnarray}
Note that $\mathcal{R}_{\mu\nu}$, $\mathcal{R}_{\mu}$, $\star$ and $D$ refer to the four-dimensional ($4$D) curvature two-form, the Ricci one-form $ \mathcal{R}_{\mu}=\iota_{ X_{\lambda}}{\mathcal{R}}^{\lambda}\,_{\mu}$, where $\iota_{ X_{\lambda}}$ is the inner product, the Hodge star operator, and the covariant derivative, respectively. The  $\{ X_{\mu}, E^{\mu}\}$ are frame and coframe fields, with duality~$E^{\nu}( X_{\mu})=\iota_{X_{\mu}} E^{\nu}=\delta_{\mu}\,^{\nu}$. Hence, the pure gravitational equations of motion ~(\ref{alter}) not only include all of the classes of the solutions to Einstein's vacuum theory (i.e., $\mathcal{R}_{\mu}=0$ and $\mathcal{R}_{\mu}=\Lambda E_{\mu}$ for all constants $\Lambda$) but also have a physical and unphysical family of solutions  with or without a spin current term (see, e.g.,~\cite{pavelle2,Dean2013}). Another physically attractive property of the considered model is the corresponding gravitational energy-momentum tensor, $\tau_{ X_{\lambda}}$,  which is symmetric, trace-free, and covariantly conservative:
\begin{equation}\label{}
 \tau_{ X_{\lambda}}\equiv\frac{1}{2}\left(\iota_{ X_{\lambda}} \mathcal{R}_{\mu\nu}\wedge{\star{ \mathcal{R}^{\mu\nu}}}- \mathcal{R}_{\mu\nu}\wedge\iota_{ X_{\lambda}}{\star{ \mathcal{R}^{\mu\nu}}}\right).
\end{equation}
This tensor is associated with the Bel$-$Robinson superenergy tensor~\cite{bel,robinson,mashhoon99}, and it has a formal analogy with Maxwell's stress-energy tensor of classical electrodynamics as well as that of the YM gauge theory~\cite{yangmills}. Furthermore, the nontrivial Lagrangian of the theory is simple, purely quadratic, and homogeneous of degree 2 in curvature, and it only includes the Kretschmann invariant~\cite{Kretschmann}. Hence, it may be thought of as the most natural model in the literature by Weyl~\cite{weyl}. We can write the 4D WY action with the coupling constant $\kappa$ as follows:
\begin{eqnarray}\label{}
 S =\int_{M_{4}}{\mathcal{L}}
=-\frac{1}{2\kappa^{2}}\int_{M_{4}} \mathcal{R}_{\mu\nu}\wedge {\star{ \mathcal{R}^{\mu\nu}}}.
\end{eqnarray}
This curvature-squared term also appears in higher-derivative gravity theories such as superstring theory~\cite{Candelas,Gross85,Gross86}, quantum gravity~\cite{lausher2002}, and  loop quantum cosmology~\cite{cognola2008}, which are perturbatively renormalized~\cite{Utiyama1,stelle1,Tomboulis1} just as the YM theories are~\cite{Slavnov,Hooft}, as well as asymptotically free~\cite{Tomboulis2,Fradkin} and capable of ameliorating ghost modes~\cite{Salam1,Stelle2,Antoniadis}.

Despite its higher-order curvature-squared action and
its system of third-order coupled nonlinear field equations,
various exact solutions to the theory have been obtained and discussed in the works of several authors. These include the static spherically symmetric solutions of Pavelle~\cite{pavelle2,pavelle1,pavelle3,pavelle4}, Thompson~\cite{thompson}, Ni~\cite{ni}, Barrent \textit{et al.}~\cite{Barrent}, Baekler \textit{et al.}~\cite{baekler1,baekler2,baekler}, and Hsu and Yeung~\cite{Hsu}; the double-dual solutions of Benn \textit{et al.}~\cite{benn81}; the anti-self-dual solutions of Mielke~\cite{mielke81}; and the pp-wave solutions of Pavelle~\cite{pavelle3}, Ba\c{s}kal~\cite{baskal}, and Kuyrukcu~\cite{kuyrukcu}. Hence, it is quite natural to discover  exact solutions  especially in more than $4$D WY theory  because  higher-dimensional models are already a standard assumption in high energy physics. Although this technique offers promising new features, finding exact solutions to the higher-dimensional models is much more difficult~\cite{Emparan}. In this scheme, we study  static spherically symmetric solutions to the non-Abelian Weyl$-$Yang$-$Kaluza$-$Klein (WYKK) gravity model~\cite{halil14} in the presence of a well-known generalized, not rotating but both electrically and magnetically charged, $SU(2)$ Wu$-$Yang monopole~\cite{wuyang,wuyang75,Actor,carmeli} and the classical $(4+3)$D non-Abelian KK metric ansatz, without loss of generality. The Wu$-$Yang monopoles (actually Wu$-$Yang solutions~\cite{Chu}) may be interpreted as non-Abelian versions of the Dirac magnetic monopole~\cite{dirac} (for a review, see, e.g.,~\cite{Goddard,Shnir}), and they are a very important gauge class for quantum chromodynamics  with instanton-like configurations~\cite{Konishi}. It should also be emphasized that the magnetic-monopole solutions (i.e., topological solitons~\cite{Preskill}) together with the dyon, instanton, and meron solutions play very significant roles, not only in $4$D Einstein$-$Yang$-$Mills$-$Higgs theories~\cite{Volkov} but also in (non)-Abelian KK theories~\cite{Witt1,sorkin,grossperry,Lee1,Perry,Angus,Cotaescu}, since they show explicitly the interaction between the gravitational, gauge, and dilatonic fields, although they are not experimentally observed. Conversely, despite being the simplest acceptable non-Abelian models~\cite{Ark}, the $6$D and $7$D theories are not as physically attractive as the $5$D Abelian and $11$D  supergravity and superstring models. However, the considered model not only contains both the graviton and the gauge boson but also the three-sphere, $S^{3}$, which being a compact $3$D maximally symmetric internal space also has advantages for $SU(2)$ gauge theories~\cite{Lunev}. Moreover, the three-sphere is simple, nontrivial, and the best-known $3$D manifold together with pseudo-sphere, and it possesses more symmetry than the other space~\cite{Okada86}.

The outline of our article is as follows. In section 2, the WY version of the non-Abelian KK theory is formulated. We investigate the reduced structures of the field equations, which include the new spin current term in terms of various combinations of two-form field strength as well as YM force density, and the stress-energy-momentum tensor, which comprises the well-known and lesser-known $4$D energy-momentum tensors of many different fields, including the WY and YM theories in the usual $4$D space-time. In section 3,  exact solutions to the third-order WY gravity model were
constructed by considering the $7$D classical non-Abelian KK metric ansatz whose $4$D part is spherically symmetric but stationary in time and whose $3$D part is the standard round and squashed three-sphere, $S^{3}$, coupled with the nonvanishing  warping function, respectively. Then, different analytic solutions are obtained. One of them is the nontrivial (anti-)dyonic black hole, which carries both unit electric and magnetic charges. It is demonstrated that the cosmological constant is just an integration constant that comes, interestingly, out of the constant warping function and the principal internal radii, which guarantee that its unit is the inverse of the square of a length. This resulting solution is also compared with the $SO(3)$ magnetic-monopole solutions of the $6$D~\cite{Perry} and $7$D~\cite{Angus} Einsteinian models, respectively. The other solutions that we found can be classified as non-Einsteinian solutions for the $J(r)\neq0$ case and two magnetic (anti-)monopole solutions, which are the $4$D and $7$D neutral Schwarzschild-like black hole solutions in the (anti-)de Sitter space-time, respectively, with the use of the Frobenius method~\cite{Arfken1985} for the $J(r)=0$ cases. It seems that obtaining a dyon/monopole or an antidyon/antimonopole solution depends on the sign of the internal space-squashing parameter, $\ell$. In this fashion, we also obtain the energy-momentum components, which are given in detail corresponding to these exact solutions. Moreover, it is shown that the existence of a nonconstant warping function is essential for finding new-type black hole solutions in the non-Abelian WYKK  gauge theory. We finally present our conclusions, remarks, and future plans in the last section.

Throughout this study,  we will apply the Maurer$-$Cartan exterior forms (which are compact-coordinate-free notations), together with geometric units (i.e., an unspecified length scale in which $c=1$, $\hbar=1$, and $8\pi G=1$) with unit coupling constants. The differential forms are known to be very useful for defining the basis of nontrivial $S^{3}$ manifolds and finding exact solutions, rather than a coordinate basis, if there are symmetry conditions on the metric tensor.

\section{Gravitational field equations and stress-energy-momentum tensor}

By considering a particular generalization of Einstein's gravity theory, the general exterior form of the Stephenson$-$Kilmister$-$Yang$-$Fairchild (SKYF) action, which is quadratic in the Riemann$-$Christofell components proportional to the real coupling constant, $\hat \kappa$, on the $(4+3)$D space-time manifold,~$(M_{7},\hat G)$ may be simply defined as follows:
\begin{eqnarray}\label{lagrange1}
\hat S [\hat{E},\hat{\Gamma}]=\int_{M_{7}}\hat{\mathcal{L}}
=\int_{M_{7}}-\frac{1}{2\hat\kappa^{2}}\hat{\mathcal{R}}_{AB}\wedge {\hat\star{\hat{\mathcal{R}}^{AB}}}+\hat\Lambda{\hat\star\mathbf{1}}+\hat{\mathcal{L}}_{m}[\hat{\Gamma}],
\end{eqnarray}
which includes a cosmological constant,~$\hat\Lambda$, and a matter-radiation field,~$\hat{\mathcal{L}}_{m}[\hat{\Gamma}]$, which is coupled to the affine connection. The quantities in the $7$D whole space (ordinary $4$D external space) are denoted by the symbols with (without) the hat symbol as a superscript. The direct product of the usual $4$D physical space-time and a $3$D internal space gives the manifold,~$M_{7}$, i.e.,~$M_{7}=M_{4}\times M_{3}$. Here, the curved subspace $M_{3}$ is taken to be the usual round three-sphere $M_{3}=\mathcal{G}/H=S^{3}$ with a radius $l=1$, and therefore, the isometry group of $S^{3}$ is defined by a special unitary group of degree $2$ as $\mathcal{G}=SU(2)\times SU(2)\simeq SO(4)$. We conveniently describe the $7$D Hodge dual operator~$\hat\star$, which is a linear map,~$ \hat\star: \mho_{p}(M_{7})\rightarrow \mho_{7-p}(M_{7})$ with respect to the Lorentzian metric tensor~$\hat G=\hat\eta_{AB}\hat E^{A}\otimes \hat E^{B}$, where~$\hat\eta_{AB}=$diag$(-1,+1,+1,+1,+1,+1,+1)$, $\mho_{p}(M_{7})$ is the vector space of p-forms,~$\hat E^{A}$ are the orthonormal basis, $\otimes$ denotes the symmetric tensor product, and the Latin capital vielbein indices~$A,B,...=0, 1, 2, 3, 5, 6, 7$ refer to the entire space,~$M_{7}$. The invariant tensor~$\hat\eta^{AB}$/~$\hat\eta_{AB}$ is used to raise/lower the indices appearing on $7$D quantities. Moreover,~$\hat\star\mathbf{1}$ is the oriented volume seven-form as~$ \hat\star\mathbf{1}={\star\mathbf{1}}\wedge\#\mathbf{1}$, where ${\star\mathbf{1}}=\hat E^{0}\wedge\hat E^{1}\wedge\hat E^{2}\wedge\hat E^{3}$ is the usual four-form invariant volume element with a $4$D Hodge map,~$\star: \mho_{p}(M_{4})\rightarrow \mho_{4-p}(M_{4})$, whereas $\#\mathbf{1}=\hat E^{5}\wedge\hat E^{6}\wedge\hat E^{7}$ is the three-form invariant volume element with a $3$D Hodge map,~$\#: \mho_{p}(S^{3})\rightarrow \mho_{3-p}(S^{3})$  with the wedge product $\wedge$.  The antisymmetric Levi$-$Civita tensor is also chosen to be~$\varepsilon_{0123567}=+1$ in this study.

Let us now investigate the coupled gravitational field equations by considering the first-order Einstein$-$Palatini variation  formalism which assumes that two fundamental variables the one-form coframes (vielbein fields),~$\hat E^{A}$, and the Levi$-$Civita connections,~$\hat\Gamma^{A}\,_{B}$ are independent of each other, and ``they are invariant under both space-time diffeomorphisms and local frame rotations,'' as noted by Daum and Reuter~\cite{daum}. Hence, we have
\begin{equation}
\delta \hat{S}[\hat{E},\hat{\Gamma}]=\int_{M_{7}}\delta\hat E^{A}\wedge\frac{\delta\hat{\mathcal{L}}}{\delta\hat E^{A}}+\delta\hat \Gamma^{AB}\wedge\frac{\delta\hat{\mathcal{L}}}{\delta\hat \Gamma^{AB}}.
\end{equation}
Therefore, we evaluate the infinitesimal variation of the generalized SKYF Lagrangian density,~$\hat{\mathcal{L}}$, in equation~(\ref{lagrange1}) by assuming that variations of the fields vanish on the boundary  as follows:
\begin{eqnarray}\label{lagrange2}
\delta \hat{S}[\hat{E},\hat{\Gamma}]&=&\int_{M_{7}}
\frac{1}{2\hat\kappa^{2}}\delta\hat E^{A}\wedge\Bigl(\hat\iota_{\hat X_{A}}\hat{\mathcal{R}}_{BC}\wedge{\hat\star{\hat{\mathcal{R}}^{BC}}}-\hat{\mathcal{R}}_{BC}\wedge\hat\iota_{\hat X_{A}}{\hat\star{\hat{\mathcal{R}}^{BC}}}\Bigl)+\nonumber\\&&
+\hat\Lambda\delta\hat E^{A}\wedge{\hat\star\hat E_{A}}
-\frac{1}{\hat\kappa^{2}}\delta\hat\Gamma^{AB}\wedge
\hat D{\hat\star{\hat{\mathcal{R}}_{AB}}}
+\delta\hat\Gamma^{AB}\wedge{\hat\star{{\hat S_{BA}}}},
\end{eqnarray}
with inner multiplication,~$ \hat\iota_{\hat X_{A}}:\mho_{p}(M_{7})\rightarrow\mho_{p-1}(M_{7})$. Here, the canonical term $\hat S_{AB}$ is the spin-density of all matter fields as a complementary term of the theory, which is determined by the variation of the matter Lagrangian $\hat{\mathcal{L}}_{m}$ with respect to the gauge potentials $\hat S_{BA}\equiv\delta\hat{\mathcal{L}}_{m}/\delta\hat \Gamma_{AB}$, and $\hat D\equiv\hat E^{A}\hat D_{\hat X_{A}}$ stands for the covariant exterior derivative. The variation principle $\delta \hat{S}=0$ gives the desired two sets of field equations. If the cosmological constant term $\hat\Lambda$ vanishes, then the first of these equations is the generalized Yang equation with the source term
\begin{eqnarray}\label{feq1}
\hat D{\hat\star{\hat{\mathcal{R}}_{AB}}}=-\hat\kappa^{2}\hat\star{\hat S_{AB}}.
\end{eqnarray}
This relationship is the master equation of the considered model with gravitational current one-form (${\hat S_{AB}}=\hat S_{ABC}\hat E^{C}$), which are antisymmetric under interchange of $A$ and $B$ indices, as expected. It is easy to verify from equation~(\ref{feq1}), and the gravitational Bianchi identities, the first $\hat{\mathcal{R}}_{AB}\wedge\hat E^{B}=0$, and the second $\hat D\hat{\mathcal{R}}_{AB}=0$, that the gauge current term  precisely  satisfies the conservation property $\hat D{\hat\star{\hat S_{AB}}}=0$ and the cyclic identity  $\hat\iota_{\hat X^{B}}\hat\iota_{\hat X^{A}}\hat\star{\hat S_{AB}}=0$, or more simply, $\hat S_{AB}\wedge\hat E^{A}\wedge\hat E^{B}=0$. Conversely, the other fundamental field equations that are obtained from the coframe variation lead to
\begin{eqnarray}\label{feq2}
\hat \tau_{\hat X_{A}}\equiv\frac{1}{2}\left(\hat\iota_{\hat X_{A}}\hat{\mathcal{R}}_{BC}\wedge{\hat\star{\hat{\mathcal{R}}^{BC}}}-\hat{\mathcal{R}}_{BC}\wedge\hat\iota_{\hat X_{A}}{\hat\star{\hat{\mathcal{R}}^{BC}}}\right),
\end{eqnarray}
where ~${\hat \tau_{\hat X_{A}}}=\hat T_{AB}{\hat\star\hat E^{B}}$ is interpreted as the canonical stress-energy-momentum tensor, which is determined by $\hat \tau_{\hat X_{A}}\equiv\hat\kappa^{2}(\delta\hat{\mathcal{L}}/\delta\hat E^{A})$. By taking the covariant exterior derivative of both sides of the energy definition~(\ref{feq2}), we obtain
\begin{eqnarray}
\hat D \hat \tau_{\hat X_{A}}=-\hat\iota_{\hat X_{A}}\hat{\mathcal{R}}_{BC}\wedge{\hat\star{\hat S^{BC}}}.
\end{eqnarray}
It can be seen above that $\hat \tau_{\hat X_{A}}$ is covariantly conservative, provided that ${\hat S_{BC}}=0$. One can also derive that the trace of the stress-energy tensor $\hat \tau_{\hat X_{A}}\wedge\hat E^{A}$ on the manifold ~$(M_{7},\hat G)$ turns to be
\begin{eqnarray}
\hat \tau_{\hat X_{A}}\wedge\hat E^{A}=-\frac{3}{2}\hat{\mathcal{R}}_{BC}\wedge{\hat\star{\hat{\mathcal{R}}^{BC}}}.
\end{eqnarray}
This is completely trace-free only in the $4$D limit.

Let us now obtain reduced forms of the field equations and stress-energy tensors in the usual 4D space-time by making use of a dimensionality reduction procedure. First, we assume that the gravitational source terms vanish and that the fields are  independent of the extracoordinates in higher dimensions.  It is convenient to also write the (4+3)D  metric tensor $\hat G(x,y)$ in the form:
\begin{eqnarray}\label{metricc1}
\hat G(x,y)=G(x)+\delta_{ij}\left(\mathcal{A}^{i}\otimes \mathcal{A}^{j}
+\mathcal{A}^{i}\otimes E^{j}+E^{i}\otimes \mathcal{A}^{j}\right)+G(y).
\end{eqnarray}
We introduce 7D coordinates as~$z^{A}=(x^{\mu},y^{i})$, where (and henceforth in this work), the Greek  vierbein indices $\mu,\nu,...=0, 1, 2, 3$ refer to the $4$D external space $M_{4}$ with the Minkowskian signature metric,~$G(x)=\eta_{\mu\nu} E^{\mu}(x)\otimes  E^{\nu}(x)$, whereas the Latin dreibein indices $i,j,...=5,6,7$ refer to the curved internal $3$D  compact subspace $S^{3}$ with the Euclidean signature metric~$G(y)=\delta_{ij}E^{i}(y)\otimes E^{j}(y)$ and~$\delta_{ij}=$diag$(+1,+1,+1)$. Additionally, YM gauge field one-form can be defined as~$\mathcal{A}^{i}=\mathcal{A}^{i}(x)=\mathcal{A}^{i}\,_{\mu}(x)E^{\mu}(x)$.

On account of the full Lorentzian metric tensor~$\hat G(x,y)=\hat\eta_{AB}\hat E^{A}\otimes \hat E^{B}$ and equation~(\ref{metricc1}), the appropriate orthonormal coframes take the simplified forms:
\begin{eqnarray}\label{coframes}
\hat E^{\mu}(x)=E^{\mu}(x),\qquad\qquad\qquad
\hat E^{i}(x,y)=\mathcal{A}^{i}(x)+E^{i}(y).
\end{eqnarray}
It follows that we immediately obtain the components of interior product operator $\hat\iota_{\hat X_{A}}$ as
\begin{eqnarray}\label{inners}
\hat\iota_{\hat X_{\mu}}(x,y)=\iota_{ X_{\mu}}-\mathcal{A}^{i}\,_{\mu}(x)\iota_{ X_{i}},\qquad\qquad\qquad
\hat\iota_{\hat X_{i}}(y)=\iota_{ X_{i}}.
\end{eqnarray}
Here, $\{\hat X_{A}\}$ are frame fields dual to the coframes~$\{\hat E^{A}\}$, together with~$\hat E^{B}(\hat X_{A})=\hat\iota_{\hat X_{A}}\hat E^{B}=\hat\delta_{A}\,^{B}$, and $\iota_{ X_{\mu}}$/$\iota_{ X_{i}}$ are the
$4$D/$3$D contraction operators, which are dual to coframes~$E^{\mu}$/$E^{i}$, i.e.,~$\iota_{ X_{\mu}}E^{\nu} = \delta_{\mu}\,^{\nu}$/$\iota_{ X_{i}}E^{j} = \delta_{i}\,^{j}$, as mentioned before. It is worthwhile to recall that the first Cartan$-$Maurer structure equation is represented by
\begin{eqnarray}\label{cartan1}
\hat D\hat E^{A}\equiv\hat d \hat E^{A}+\hat\Gamma^{A}\,_{B}\wedge\hat E^{B}=\hat T^{A},
\end{eqnarray}
with the seven-exterior derivative~$\hat d\equiv\hat E^{A}\hat d_{\hat X_{A}}$. Equation~(\ref{cartan1}) can be solved for the constraint equation~$\hat T^{A}\equiv (1/2)\hat T_{BC}\,^{A}\hat E^{B}\wedge\hat E^{C}=0$, as follows:
\begin{eqnarray}\label{connectionsshort}
\hat\Gamma_{AB}=\frac{1}{2}\left[\hat\iota_{\hat X_{B}}\hat d \hat E_{A}-
\hat\iota_{\hat X_{A}}\hat d \hat E_{B}+(\hat\iota_{\hat X_{A}}\hat\iota_{\hat X_{B}}\hat d \hat E_{C})\hat E^{C}\right],
\end{eqnarray}
where $\hat\Gamma_{AB}$ are the one-form Levi$-$Civita connections, i.e., $\hat\Gamma_{(AB)}=0$, which means that the metric compatibility condition~$\hat D\hat G(x,y)=0$ is valid for the torsion-free case~\cite{benntucker}. Hence, substituting equations~(\ref{coframes}) and~(\ref{inners}) into the equation~(\ref{connectionsshort}), we directly obtain the following components of the spin connection $\hat\Gamma_{AB}$:
\begin{eqnarray}\label{connections}
\fl\hat\Gamma^{\mu}\,_{\nu}=\Gamma^{\mu}\,_{\nu}-\frac{1}{2} F^{i\mu}\,_{\nu}\hat E_{i},\quad~~
\hat\Gamma_{\mu}\,^{j}=-\hat\Gamma^{j}\,_{\mu}=-\frac{1}{2}F^{j}\,_{\mu\nu}E^{\nu},\quad~~
\hat\Gamma^{i}\,_{j}=\Gamma^{i}\,_{j}-\frac{1}{2}\varepsilon^{i}\,_{jk}\mathcal{A}^{k}.
\end{eqnarray}
By assuming that the $T_{i}$ are the generators of the group $SU(\mathcal{N})$, the corresponding Lie algebra is usually defined as $[T_{i},T_{j}]=f^{k}\,_{ij}T_{k}$ with real structure constants $f^{k}\,_{ij}$ and $T_{i}= (1/2)\sigma_{i}$, where $\sigma_{i}$ are the Pauli matrices. We also have $f^{k}\,_{ij}=\varepsilon^{k}\,_{ij}$ for the gauge group $SU(2)$ and the little group $SO(3)$.  Hence, we can prove that the coframes of internal $S^{3}$ space satisfy the right-invariant structure equations (see, e.g., ~\cite{Lee83})
\begin{eqnarray}\label{rightinv}
dE^{i}(y)=\frac{1}{2}\varepsilon^{i}\,_{jk}E^{j}(y)\wedge E^{k}(y),
\end{eqnarray}
meaning that $\Gamma^{i}\,_{j}=(1/2)\varepsilon^{i}\,_{jk}E^{k}$. Moreover, the $SU(2)$-valued  two-form field strength is explicitly determined by
~$F^{i}\equiv d\mathcal{A}^{i}+(1/2)e\varepsilon^{i}\,_{jk}\mathcal{A}^{j}\wedge \mathcal{A}^{k}=(1/2)F^{i}\,_{\mu\nu}E^{\mu}\wedge E^{\nu}$. Here, $e$ is nothing but the coupling constant, and for convenience, it can be taken as unity, $e=1$.

We now consider the second Maurer$-$Cartan structure equation,
\begin{eqnarray}\label{cartan2}
\hat D \hat\Gamma^{A}\,_{B}\equiv\hat d\hat \Gamma^{A}\,_{B}+\hat\Gamma^{A}\,_{C}\wedge\hat\Gamma^{C}\,_{B}=\hat{\mathcal{R}}^{A}\,_{B},
\end{eqnarray}
to obtain the dimensionally-reduced curvature two-form $\hat{\mathcal{R}}^{A}\,_{B}$, where $\hat{\mathcal{R}}^{A}\,_{B}\equiv (1/2)\hat R^{A}\,_{BCD}\hat E^{C}\wedge\hat E^{D}$. By substituting gauge potentials one-form~(\ref{connections}) into equation~(\ref{cartan2}) and performing some algebra, the first component,~$\hat{\mathcal{R}}_{\mu\nu}$, becomes (see, e.g., ~\cite{dereli90})
\begin{eqnarray}\label{curvature1}
\hat{\mathcal{R}}^{\mu}\,_{\nu}=\Psi^{\mu}\,_{\nu}+\Omega^{\mu}\,_{\nu}+\Sigma^{\mu}\,_{\nu},
\end{eqnarray}
where the two-form terms are
\begin{eqnarray}\label{connection1s}
\Psi^{\mu}\,_{\nu}&=&\mathcal{R}^{\mu}\,_{\nu}-\frac{1}{2}F^{i\mu}\,_{\nu}F_{i}-\frac{1}{4}
\iota_{ X^{\mu}}F^{i}\wedge \iota_{ X_{\nu}}F_{i},\nonumber\\
\Omega^{\mu}\,_{\nu}&\equiv&\Omega^{i\mu}\,_{\nu}\wedge\hat E_{i}=-\frac{1}{2}\mathcal{D}F^{i\mu}\,_{\nu}\wedge\hat E_{i},\nonumber\\
\Sigma^{\mu}\,_{\nu}&\equiv&\Sigma^{ij\mu}\,_{\nu}\hat E_{i}\wedge\hat E_{j}
=\frac{1}{4}(F^{i\mu\lambda}F^{j}\,_{\lambda\nu}-\varepsilon^{ij}\,_{k}
F^{k\mu}\,_{\nu})\hat E_{i}\wedge\hat E_{j}.
\end{eqnarray}
Here, we also make use of the ordinary curvature two-form,~$\mathcal{R}_{\mu\nu}$, of the external space, $M_{4}$,  and the YM covariant derivative,~$\mathcal{D}F^{i}\,_{\mu\nu}=D F^{i}\,_{\mu\nu}+e\varepsilon^{i}\,_{jk}\mathcal{A}^{j}F^{k}\,_{\mu\nu}$ in the adjoint representation. It follows that the remaining nonvanishing components can be written as
\begin{eqnarray}\label{curvature2}
\fl\hat{\mathcal{R}}_{\mu }\,^{j}=-\hat{\mathcal{R}}^{j}\,_{\mu }=\Omega^{j}\,_{\mu\nu}\wedge E^{\nu}-\Sigma^{ij}\,_{\mu\nu} E^{\nu}\wedge\hat E_{i},\quad\quad
\hat{\mathcal{R}}^{i}\,_{j}=\Sigma^{i}\,_{j\mu\nu} E^{\mu}\wedge E^{\nu}+\frac{1}{4}\hat E^{i}\wedge\hat E_{j}.\label{curvature3}
\end{eqnarray}
It is not difficult to conjecture that, in order to find the desired field equations and stress-energy tensors, we need not only the curvature two-form~(\ref{curvature1})--(\ref{curvature3}) but also the reduced forms of various Hodge duality identities, which are found to be
\begin{eqnarray}\label{hodge}
\fl{\hat\star{\hat E^{\mu}}}={\star{E^{\mu}}}\wedge\#1,\qquad
{\hat\star{\hat E^{i}}}=\# E^{i}\wedge{\star1},\qquad
{\hat\star({\hat E^{\mu}}\wedge{\hat E^{\nu}})}={\star({E^{\mu}}\wedge{ E^{\nu}})}\wedge\#1,
\nonumber\\
\fl{\hat\star({\hat E^{\mu}}\wedge{\hat E^{j}})}
=-{\hat\star({\hat E^{j}}\wedge{\hat E^{\mu}})}
=-{\star{E^{\mu}}}\wedge\# E^{j},
\qquad
{\hat\star({\hat E^{i}}\wedge{\hat E^{j}})}=\#({E^{i}}\wedge{ E^{j}})\wedge{\star1}.
\end{eqnarray}

\subsection{Reduced field equations}

Taking into account the definition of a covariant exterior derivative, the second gravitational Bianchi identity is manifestly written as
\begin{eqnarray}\label{}
\hat D{{\hat{\mathcal{R}}^{A}\,_{B}}}=\hat d{{\hat{\mathcal{R}}^{A}\,_{B}}}
+\hat \Gamma^{A}\,_{C}\wedge{{\hat{\mathcal{R}}^{C}\,_{B}}}
-\hat \Gamma^{C}\,_{B}\wedge{{\hat{\mathcal{R}}^{A}\,_{C}}}=0,
\end{eqnarray}
enabling  us to arrive at the source-free field equations~(\ref{feq1}) in the following form:
\begin{eqnarray}\label{}
\hat D{\hat\star{\hat{\mathcal{R}}^{A}\,_{B}}}=\hat d{\hat\star{\hat{\mathcal{R}}^{A}\,_{B}}}
+\hat \Gamma^{A}\,_{C}\wedge{\hat\star{\hat{\mathcal{R}}^{C}\,_{B}}}
-\hat \Gamma^{C}\,_{B}\wedge{\hat\star{\hat{\mathcal{R}}^{A}\,_{C}}}=0.
\end{eqnarray}

After some lengthy but tedious calculations and remembering that the antisymmetry property of the wedge product (${\hat E^{A}}\wedge{\hat E^{B}}\equiv{\hat E^{A}}\otimes{\hat E^{B}}-{\hat E^{B}}\otimes{\hat E^{A}}$) gives some extra terms, we obtain six reduced equations including the gravitational and the gauge fields, which are determined using three field equations,
$\hat D{\hat\star{\hat{\mathcal{R}}^{\mu}\,_{v}}}=0$, $\hat D{\hat\star{\hat{\mathcal{R}}^{\mu}\,_{j}}}=0$, and $\hat D{\hat\star{\hat{\mathcal{R}}^{i}\,_{j}}}=0$. Their final forms are explicitly written in a Cartesian chart as follows:
\begin{eqnarray}
\fl D_{\lambda}{{ R^{\lambda}\,_{\sigma\mu\nu}}}=S_{\mu\nu\sigma},\label{a4}\\
\fl\mathcal{D}_{\lambda}\mathcal{D}^{\lambda}F^{i}\,_{\mu\nu}\label{rfe2}
+R_{\mu\nu\lambda\rho}F^{i\lambda\rho}
+F^{i}\,_{\mu\nu}
-\frac{1}{2}(F^{i}\,_{\lambda\rho}F^{j\lambda\rho}F_{j\mu\nu}
-F^{i}\,_{\mu\lambda}F^{j\lambda\rho}F_{j\rho\nu}+\nonumber\\
\fl+F^{i}\,_{\nu\lambda}F^{j\lambda\rho}F_{j\rho\mu})
+\varepsilon^{i}\,_{kj} (F^{k}\,_{\mu\lambda}F^{j\lambda}\,_{\nu}-F^{j}\,_{\mu\lambda}F^{k\lambda}\,_{\nu})=0,\\
\fl\mathcal{D}_{\lambda}\mathcal{D}_{\mu}F^{i\lambda}\,_{\sigma}\label{rfe3}
+R_{\mu\lambda\sigma\rho}F^{i\lambda\rho}
+\frac{1}{2}F^{i}\,_{\mu\sigma}
-\frac{1}{4}(F^{i}\,_{\lambda\rho}F^{j\lambda\rho}F_{j\mu\sigma}
+2F^{i}\,_{\sigma\lambda}F^{j\lambda\rho}F_{j\rho\mu})-\nonumber\\
\fl-\varepsilon^{i}\,_{jk} F^{j}\,_{\mu\lambda}F^{k\lambda}\,_{\sigma}=0,\\
\fl\mathcal{D}_{\mu}(F^{i}\,_{\lambda\rho}F^{j\lambda\rho})\label{rfe4}
+F^{i}\,_{\mu}\,^{\lambda}J^{j}\,_{\lambda}
-\varepsilon^{ij}\,_{k} J^{k}\,_{\mu}=0,\\
\fl F^{j\lambda}\,_{\mu}J^{i}\,_{\lambda}-F^{i\lambda}\,_{\mu}J^{j}\,_{\lambda}
 -2\varepsilon^{ij}\,_{k} J^{k}\,_{\mu}=0,\label{rfe5}\\
\fl\varepsilon^{ik}\,_{l}F^{j}\,_{\lambda\rho}F^{l\lambda\rho}\label{rfe6}
-\varepsilon^{jk}\,_{l}F^{i}\,_{\lambda\rho}F^{l\lambda\rho}
+2\varepsilon^{ij}\,_{l}F^{k}\,_{\lambda\rho}F^{l\lambda\rho}=0,
\end{eqnarray}
where $J^{i}\,_{\nu}=\mathcal{D}_{\mu}F^{i\mu}\,_{\nu}$ is the YM current one-form (i.e., $\mathcal{D}\star{F^{i}}=\star{J^{i}}$) and the right side of equation~(\ref{a4}) can be described as a 4D source term of the matter field
\begin{eqnarray}\label{rfe16}
\fl S_{\mu\nu\sigma}=\frac{1}{2}\biggl[ F_{i\mu\nu}J^{i}\,_{\sigma}
 +\frac{1}{2}(F_{i\sigma\nu}J^{i}\,_{\mu}-F_{i\sigma\mu}J^{i}\,_{\nu})
+\mathcal{D}_{\nu}(F_{i\mu}\,^{\lambda}F^{i}\,_{\lambda\sigma})
-\mathcal{D}_{\mu}(F_{i\nu}\,^{\lambda}F^{i}\,_{\lambda\sigma})
\biggl].
\end{eqnarray}
By considering the equation~(\ref{alter}) and  the following identity
\begin{eqnarray}
[\mathcal{D}_{\mu},\mathcal{D}_{\nu}]F^{i\lambda}\,_{\sigma}
=R^{\lambda}\,_{\tau\mu\nu}F^{i \tau}\,_{\sigma}-R^{\tau}\,_{\sigma \mu\nu}F^{i\lambda}\,_{\tau}
+\varepsilon^{i}\,_{jk}F^{j}\,_{\mu\nu}F^{k\lambda}\,_{\sigma},
\end{eqnarray}
the reduced field equations~(\ref{a4})--(\ref{rfe6}) can be written in terms of the reduced components of the one-form Ricci tensor ${\hat{\mathcal{R}}}_{A}$, ${\hat{\mathcal{R}}}_{A}=\hat\iota_{\hat X_{B}}{\hat{\mathcal{R}}}^{B}\,_{A}=\hat R_{AB}\hat E^{B}$,
\begin{eqnarray}\label{Einsteinian}
\hat R_{\mu\nu}\equiv\mathcal{P}_{\mu\nu}= R_{\mu\nu}-\frac{1}{2}F^{i}\,_{\mu\lambda}F_{i\nu}\,^{\lambda},\nonumber\\
\hat R_{i\nu}=\hat R_{\nu i}\equiv\mathcal{Q}_{i\nu} = \mathcal{D}_{\mu}F_{i}\,^{\mu}\,_{\nu},\nonumber\\
\hat R_{ij}\equiv\mathcal{U}_{ij} = F_{i\mu\nu}F_{j}\,^{\mu\nu}+\varepsilon_{ikl}\varepsilon_{j}\,^{kl},
\end{eqnarray}
with $\varepsilon_{ikl}\varepsilon_{j}\,^{kl}=2\delta_{ij}$ for the 3D internal space, as follows:
\begin{eqnarray}
&&\mathcal{D}_{\mu}\mathcal{P}_{\sigma\nu}
-\mathcal{D}_{\nu}\mathcal{P}_{\sigma\mu}
+\frac{1}{4}(F^{i}\,_{\sigma\mu}\mathcal{Q}_{i\nu}
-F^{i}\,_{\sigma\nu}\mathcal{Q}_{i\mu}
-2F^{i}\,_{\mu\nu}\mathcal{Q}_{i\sigma})=0,\label{emb1}\\
&&\mathcal{D}_{\mu}\mathcal{Q}_{i\nu}
-\mathcal{D}_{\nu}\mathcal{Q}_{i\mu}
+F_{i}\,^{\lambda}\,_{\nu}\mathcal{P}_{\lambda\mu}
-F_{i}\,^{\lambda}\,_{\mu}\mathcal{P}_{\lambda\nu}
+\frac{1}{2}F^{j}\,_{\mu\nu}\mathcal{U}_{ij}=0,\label{cfe1}\\
&&\mathcal{D}_{\mu}\mathcal{Q}_{i\sigma}
+F_{i}\,^{\lambda}\,_{\sigma}\mathcal{P}_{\lambda\mu}
+\frac{1}{4}F^{j}\,_{\sigma\mu}\mathcal{U}_{ij}=0,\label{cfe2}\\
&&\mathcal{D}_{\mu}\mathcal{U}_{ij}
+F_{i\mu}\,^{\lambda}\mathcal{Q}_{j\lambda}
-\varepsilon_{ij}\,^{k}\mathcal{Q}_{k\mu}
=0,\label{emb3}\\
&&F_{j}\,^{\lambda}\,_{\mu}\mathcal{Q}_{i\lambda }-F_{i}\,^{\lambda}\,_{\mu}\mathcal{Q}_{j\lambda}
 -2\varepsilon_{ij}\,^{k} \mathcal{Q}_{k\mu}=0 ,\label{emb4}\\
&&\varepsilon_{ik}\,^{l}\mathcal{U}_{jl}
-\varepsilon_{jk}\,^{l}\mathcal{U}_{il}
+2\varepsilon_{ij}\,^{l}\mathcal{U}_{kl}=0 \label{emb2}.
\end{eqnarray}
One can see that equations~(\ref{a4})--(\ref{rfe6}) and~(\ref{emb1})--(\ref{emb2}) possess mathematically a richer structure than those found in Einstein's theory (\ref{Einsteinian}). There is no doubt that the equations ~(\ref{emb1})--(\ref{emb2}) naturally contain the Einsteinian solutions ${\hat{\mathcal{R}}}_{A}=0$,
\begin{equation}\label{n19}
{\hat{\mathcal{R}}}_{A}=0 \quad\Rightarrow\quad
\cases{
\mathcal{P}_{\mu\nu}=0 & \quad $\Rightarrow$ \quad $R_{\mu\nu}=\frac{1}{2}F^{i}\,_{\mu\lambda}F_{i\nu}\,^{\lambda},$\\
\mathcal{Q}_{i\nu}=0 & \quad $\Rightarrow$ \quad $\mathcal{D}_{\mu}F_{i}\,^{\mu}\,_{\nu}=0,$\\
~\mathcal{U}_{ij}=0 & \quad $\Rightarrow$ \quad $F_{i\mu\nu}F_{j}\,^{\mu\nu}
=2\delta_{ij}.$\\}
\end{equation}
From equations~(\ref{rfe4}) and~(\ref{rfe5}), we also have
\begin{eqnarray}\label{lfd}
f^{i}\,_{\mu j}+f_{j\mu }\,^{i}=-4\mathcal{D}_{\mu}(F^{i}\wedge \star{F_{j}}),
\end{eqnarray}
which means that the sum of the Lorentz force density-like terms are equal to the negative gradient of the $F^{i}\wedge \star{F_{j}}$ term, which is a well-known invariant of non-Abelian theory \cite{halil13,halil14}. The four-form Lorentz force density, $f^{i}\,_{\mu i}=-\iota_{X_{\mu}}F^{i}\wedge \star{J_{i}}$, which usually appears in the geodesic equations of non-Abelian KK theories (see, e.g.,~\cite{kerner,orzalesi}) comes,
interestingly, out of the third-order field equations. Moreover, the usual $4$D continuity equation,~$\mathcal{D}\star{J^{i}}=0$, is easily obtained with the help of the equations~(\ref{rfe2}) and~(\ref{rfe3}).

\subsection{Reduced stress-energy-momentum tensor}

We can also obtain dimensionally reduced components of the energy-momentum tensor of the matter fields~$\hat \tau_{\hat X_{A}}$ by considering equation~(\ref{feq2}) together with the equations~(\ref{curvature1})--(\ref{hodge}). These equations are not only rather lengthy and complicated but also include the well-known WY energy-momentum tensor
\begin{eqnarray}
\tau^{{WY}}_{X_{\mu}}=
\frac{1}{2}\left( \iota_{X_{\mu}} \mathcal{R}_{\nu\lambda}\wedge{\star{ \mathcal{R}^{\nu\lambda}}}
- \mathcal{R}_{\nu\lambda}\wedge\iota_{X_{\mu}}{\star{ \mathcal{R}^{\nu\lambda}}}\right),
\end{eqnarray}
and that of YM theory
\begin{eqnarray}
\tau^{{YM}}_{X_{\mu}}=
\frac{1}{2}\left(\iota_{X_{\mu}} F_{i}\wedge{\star{ F^{i}}}
- F_{i}\wedge\iota_{X_{\mu}}{\star{ F^{i}}}\right),
\end{eqnarray}
as well as those of the various constituent fields, such as those which are nonminimal, cubic and quartic. For completeness, we explicitly list here the reduced components of the energy-momentum six-form of $\hat \tau_{\hat X_{A}}$ by  virtue of~${\hat \tau_{\hat X_{\mu}}}=\hat T_{\mu\nu}{\hat\star\hat E^{\nu}}+\hat T_{\mu j}{\hat\star\hat E^{j}}$ and~${\hat \tau_{\hat X_{i}}}=\hat T_{ i\nu}{\hat\star\hat E^{\nu}}+\hat T_{ij}{\hat\star\hat E^{j}}$ in the $(4+3)$ decomposition. We can also check by applying the components below whether the full energy-momentum tensor, $\hat{T}_{AB}$, satisfies the conservation property $\hat D_{A}\hat{T}^{A}\,_{B}=0$ or not. Then, direct calculations give the first component of the energy tensor as follows:
\begin{eqnarray}\label{tfse11}
\fl\hat{T}_{\mu\nu}=R_{\mu\lambda\sigma\rho}
R_{\nu}\,^{\lambda\sigma\rho}
-\frac{1}{4}\eta_{\mu\nu}R_{\tau\lambda\sigma\rho}R^{\tau\lambda\sigma\rho}
+\frac{3}{2}\left(F_{i\mu\lambda}
F^{i}\,_{\nu}\,^{\lambda}
-\frac{1}{4}\eta_{\mu\nu}F_{i\tau\lambda}
F^{i\tau\lambda}\right)+
\nonumber \\
\fl\quad\quad+\frac{3}{2}
F^{i\rho\sigma}\left[ F_{i(\mu}\,^{\lambda}R_{\nu)\lambda\sigma\rho}
-\frac{1}{4}\eta_{\mu\nu}F_{i}\,^{\tau\lambda}R_{\tau\lambda\sigma\rho}\right]+
\nonumber \\
\fl\quad\quad+\frac{3}{2}\varepsilon_{ijk}
\Big[\frac{1}{2}F^{i}_{(\mu|\sigma}F^{j\sigma\rho}F^{k}\,_{\rho|\nu)}
-\frac{1}{4}\eta_{\mu\nu}F^{i}\,_{\tau\sigma}F^{j\sigma\rho}F^{k}\,_{\rho}\,^{\tau}\Big]+\nonumber \\
\fl\quad\quad
+\frac{1}{4}\left(F_{i\mu\lambda}F_{j}\,^{\lambda\sigma}
F^{i}\,_{\sigma\rho}F^{j\rho}\,_{\nu}
-\eta_{\mu\nu}F_{i\tau\lambda}F_{j}\,^{\lambda\sigma}F^{i}\,_{\sigma\rho}
F^{j\rho\tau}\right)+\nonumber \\
\fl\quad\quad
+\frac{3}{8}F^{i}\,_{\sigma\rho}F^{j\sigma\rho}
\left(F_{i\mu\lambda}F_{j\nu}\,^{\lambda}-\frac{1}{4}\eta_{\mu\nu}F_{i\tau\lambda}
F_{j}\,^{\tau\lambda}\right)+
\nonumber
\\
\fl\quad\quad+\frac{1}{4}
\left(F_{i\mu\lambda}F_{j}\,^{\lambda\sigma}F^{j}\,_{\sigma\rho}
F^{i\rho}\,_{\nu}
-\eta_{\mu\nu}F_{i\tau\lambda}F_{j}\,^{\lambda\sigma}
F^{j}\,_{\sigma\rho}F^{i\rho\tau}\right)
-\frac{1}{16}\eta_{\mu\nu}\delta_{ij}\delta^{ij}+
\nonumber
\\
\fl\quad\quad+\left(\frac{1}{2}{\mathcal{D}}_{\sigma}F_{i\mu\rho}
{\mathcal{D}}^{\sigma}F^{i}\,_{\nu}\,^{\rho}
+\frac{1}{4}{\mathcal{D}}_{\mu}F_{i\sigma\rho}
{\mathcal{D}}_{\nu}F^{i\sigma\rho}
-\frac{1}{4}\eta_{\mu\nu}{\mathcal{D}}_{\sigma}F_{i\tau\rho}
{\mathcal{D}}^{\sigma}F^{i\tau\rho}\right).
\end{eqnarray}
It should be emphasized here that the $\eta_{\mu\nu}$ and $\delta_{ij}$ are the vierbein and dreibein metrics of the external and internal manifolds, respectively, and that $\delta_{ij}\delta^{ij}=3$. The remaining components correspondingly are given by
\begin{eqnarray}\label{se22}
\fl\hat{T}_{\mu j}=\frac{1}{2}R_{\mu\lambda\sigma\rho}
{\mathcal{D}}^{\lambda}F_{j}\,^{\sigma\rho}
+\frac{1}{2}\varepsilon_{jik}F^{i\sigma\rho}
{\mathcal{D}}_{(\mu}F^{k}\,_{\rho)\sigma}
+\frac{3}{4}F_{i\mu}\,^{\lambda}F^{i\sigma\rho}
{\mathcal{D}}_{\sigma}F_{j\rho\lambda}-\nonumber \\
\fl\quad\quad-\frac{1}{2}F_{j}\,^{\sigma\lambda}F_{i\lambda}\,^{\rho}
{\mathcal{D}}_{(\mu}F^{i}\,_{\sigma)\rho},\\
\fl\hat{T}_{ij}=
\frac{3}{8}
\left[\varepsilon_{kli}F^{l}\,_{\tau\sigma}F_{j}\,^{\sigma\rho}F^{k}\,_{\rho}\,^{\tau}
+\varepsilon_{klj}F^{l}\,_{\tau\sigma}F_{i}\,^{\sigma\rho}F^{k}\,_{\rho}\,^{\tau}
-\delta_{ij}\varepsilon_{klm}F^{l}\,_{\tau\sigma}F^{m\sigma\rho}
F^{k}\,_{\rho}\,^{\tau}\right]-
\nonumber \\
\fl\quad\quad-\frac{1}{16}\left(\frac{1}{2}F_{i\tau\lambda}F_{k}\,^{\lambda\sigma}
F_{j\sigma\rho}F^{k\rho\tau}
+\delta_{ij}F_{l\tau\lambda}F_{k}\,^{\lambda\sigma}
F^{l}\,_{\sigma\rho}F^{k\rho\tau}\right)
+\nonumber
 \\
 \fl\quad\quad+\frac{1}{4}\left(\frac{5}{2}F_{i\tau\lambda}F_{k}\,^{\lambda\sigma}
F^{k}\,_{\sigma\rho}F_{j}\,^{\rho\tau}
-\delta_{ij}F_{l\tau\lambda}F_{k}\,^{\lambda\sigma}
F^{k}\,_{\sigma\rho}F^{l\rho\tau}\right)
-\nonumber \\
\fl\quad\quad-\frac{1}{4}\delta_{ij}\left(R_{\mu\nu\lambda\rho}R^{\mu\nu\lambda\rho}
 -\frac{3}{2} F^{k\mu\nu}F_{k}\,^{\lambda\rho}R_{\mu\nu\lambda\rho}
 +\frac{3}{8}F_{k\mu\nu}F^{k\mu\nu}F_{l\lambda\rho}F^{l\lambda\rho}
-\frac{1}{4}\right)+\nonumber \\
\fl\quad\quad
+\frac{1}{4}\left({\mathcal{D}}_{\sigma}F_{i\tau\rho}
{\mathcal{D}}^{\sigma}F_{j}\,^{\tau\rho}
-\delta_{ij}{\mathcal{D}}_{\sigma}F_{k\tau\rho}
{\mathcal{D}}^{\sigma}F^{k\tau\rho}\right)
-\frac{3}{8}F_{i\lambda\rho}F_{j}\,^{\lambda\rho}.
\end{eqnarray}
It is not difficult to conjecture that for the $4$D case and $\mathcal{A}^{i}\,_{\mu}=0$, we only obtain the source-free field equation  and the stress-energy tensor of the $4$D pure WY theory, as follows:
\begin{eqnarray}
\fl D{\star{ \mathcal{R}_{\mu\nu}}}=0,\qquad\qquad\qquad
\tau^{{WY}}_{X_{\mu}}=
\frac{1}{2}\left( \iota_{X_{\mu}} \mathcal{R}_{\nu\lambda}\wedge{\star{ \mathcal{R}^{\nu\lambda}}}
- \mathcal{R}_{\nu\lambda}\wedge\iota_{X_{\mu}}{\star{ \mathcal{R}^{\nu\lambda}}}\right).
\end{eqnarray}

\section{Ansatz and static black hole solutions}

Now, we investigate the black hole solutions to the $(4+3)$D WY gravity model by considering the generalized Wu$-$Yang ansatz (for a review of higher-dimensional black hole solutions for different manners, see, e.g., ~\cite{Emparan}). It is very difficult to obtain the exact solutions of higher-derivative gravity theories.  To obtain the computational advantage in solving third-order nonlinear partial differential equations, it is useful to consider the symmetric version of the $4$D static spherically symmetric space-time metric, which  describes  not only charged but also asymptotically (anti-)de Sitter black holes. In the usual Schwarzschild coordinates, $x=(t,r,\theta,\phi)$, the external metric can put into the form:
\begin{eqnarray}\label{4metric}
\fl G(x)=-f(r)dt\otimes dt+f(r)^{-1}dr\otimes dr
+r^{2}d\theta\otimes d\theta
+r^{2}\sin^{2}\theta d\phi\otimes d\phi,
\end{eqnarray}
where the metric function $f(r)$ depends only on the radial coordinate $r$. Moreover, we choose the $3$D line element for the internal space as the unit three-sphere $S^{3}$  in terms of Euler coordinates, $y=(\Theta,\Phi,\psi)$~\cite{Henderson}, describing rotations, which are given by the line element in its standard form:
\begin{eqnarray}\label{3metric}
\fl  G(y)=d\Theta\otimes d\Theta+\sin^{2}\Theta d\Phi\otimes d\Phi
+(d\psi+\cos\Theta d\Phi)\otimes (d\psi+\cos\Theta d\Phi).
\end{eqnarray}
Here, the ranges are $0\leq \theta,\Theta \leq \pi$, $0\leq \phi,\Phi \leq 2\pi$,  $0\leq \psi \leq 4\pi$, and $0<r<r_{\infty}$, and $\partial/\partial\Phi=\partial_{\Phi}$ is the usual Killing field associated with the circle $S^{1}$ in the background of the Abelian KK theories. The internal metric  $G(y)$ in equation (\ref{3metric}) is actually a special case of Berger sphere~\cite{Berger61}, and it can be rewritten in the general rescaled, i.e., ``squashed" form:
\begin{eqnarray}\label{3metric2}
G(y)=\sum_{i=5}^{7}l^{2}_{i}E^{i}(y)\otimes E^{i}(y),
\end{eqnarray}
where the $E^{i}$ are the right-invariant one-form set by
\begin{eqnarray}
 E^{5}(y)&=&\cos\psi d\Theta+\sin\psi\sin\Theta d\Phi,\nonumber\\
E^{6}(y)&=&-\sin\psi d\Theta+\cos\psi\sin\Theta d\Phi,\nonumber\\
 E^{7}(y)&=&d\psi +\cos\Theta d\Phi.
\end{eqnarray}
Other choices of $E^{i}$ that satisfy the structure equation~(\ref{rightinv}) are also possible (see, e.g., ~\cite{Taub51,Newman63,Misner63}). The $l_{i}$ can be defined as the curvature radii of the internal space, and they are constants for a static background. We consider the symmetric case where the three principal radii of curvature are equal, $l_{5}=l_{6}=l_{7}$, corresponding to the usual round three-sphere with isometry group $\mathcal{G}=SU(2)\times SU(2)\simeq SO(4)$~\cite{Shen87}. In that special case, the $S^{3}$ manifold is completely invariant under the right action of the gauge group, $SU(2)$~\cite{Duff}. We also investigate the case where $l_{5}=l_{6}\neq l_{7}$, which corresponds to the squashed sphere geometry with isometry algebra $\mathcal{G}=SU(2)\times U(1)$. It possesses a circular isometry, and it usually appears in the quantum field theory, especially in the mixmaster Universe (see ~\cite{Dowker} and references therein). Here, it can be useful to consider the zero-form warping  function, $\varphi=\varphi(r)$, to investigate the special effects of this field on the nontrivial solutions. Hence, the final form of the full metric $\hat G(x,y)$ becomes
\begin{eqnarray}\label{metricc2}
\hat G(x,y)=G(x)+\varphi^{2}(r)\sum_{i=5}^{7}\left[l_{i}E^{i}(y)+\mathcal{A}^{i}(x)\right]\otimes \left[l_{i}E^{i}(y)+\mathcal{A}^{i}(x)\right].
\end{eqnarray}

We also introduce well-known, time-independent, generalized Wu$-$Yang monopole-like gauge potentials. Hence, in general, the gauge field ansatz, $\mathcal{A}^{i}(x)=A^{\alpha}(x)$ where $\alpha=i-4$, is given by the following spherically symmetric electric and magnetic parts:
\begin{eqnarray}\label{WuYangansatz}
A^{\alpha}(x)={A}^{\alpha}_{0}(\vec{x},t)dt+{A}^{\alpha}_{a}(\vec{x},t)dx^{a},
\end{eqnarray}
where
\begin{eqnarray}
{A}^{\alpha}_{0}(\vec{x},t)=\frac{J(r)}{g r}\hat{x}^{\alpha}, \qquad\qquad\qquad
{A}^{\alpha}_{a}(\vec{x},t)=\varepsilon^{\alpha}\,_{ab}\frac{[K(r)-1]}{g r}\hat{x}^{b},\label{gauge2}
\end{eqnarray}
which are free from any line singularity if we assume all YM fields to be independent of the extra internal coordinates. The color indices become $\alpha,\beta,...=1,2,3$ (if $i,j,...=5,6,7$), and  $a,b,...=1,2,3$ are also the external spatial indices, which mean that both indices are used in the same way. Here, the profile functions, $J(r)$ and $K(r)$, are arbitrary real-valued functions depending on the variable $r=|\vec{x}|$, to be determined by the equations of motion. As usual, $g$ is the quantized monopole charge and $\hat{x}^{a}=x^{a}/|\vec{x}|$ are the external spatial unit vectors.  After some manipulations in Minkowski space-time, the corresponding  YM field strength two-form, in which the electrically and magnetically charged but-not-rotating particles moving  with unit charge and coupling constant (i.e., $g=e=1$), can explicitly be obtained as follows:
\begin{eqnarray}\label{YMF}
F^{\alpha}(x)&=&\left\{\delta^{\alpha}_{a}\left(\frac{JK}{r^{2}}\right)
+{x}^{\alpha}{x}_{a}\left[\frac{rJ'-J(K+1)}{r^{4}}\right]\right\}
dx^{a}\wedge dt+\nonumber\\&&
+\varepsilon_{ab}\,^{c}\left[\delta^{\alpha}_{c}\left(\frac{K'}{r}\right)
-\frac{{x}^{\alpha}{x}_{c}}{r^{3}}
\left(K'-\frac{K^{2}-1}{r}\right)
\right]dx^{a}\wedge dx^{b},
\end{eqnarray}
(see also, e.g.,~\cite{dereli93,Ngome}). Here, prime implies a derivative taken with respect to $r$, as usual. If we define the electric-like field one-form as $E^{\alpha}\equiv E^{\alpha}_{a}dx^{a}$ and the magnetic-like field two-form as $B^{\alpha}\equiv B^{\alpha}_{a}\star(dx^{a}\wedge dt)$, then we can write $F^{\alpha}=E^{\alpha}\wedge dt+B^{\alpha}$. Consequently, the electric fields and the magnetic fields can be obtained using equation (\ref{YMF}) by considering the relationships $E^{\alpha}\equiv \iota_{X_{t}}F^{\alpha}$ and $B^{\alpha}\equiv -\iota_{X_{t}}\star F^{\alpha}$~\cite{benntucker1}.

By substituting the squashed version of the (4+3)D metric tensor (\ref{metricc2}) where $l_{5}=l_{6}=l$ and $l_{7}=\ell$, together with the gauge potentials (\ref{gauge2}), into the master field equation (\ref{feq1}) (${\hat S_{AB}}=0$), we unfortunately obtain complicated and highly non-linear sets of field equations because of the nature of the WY model (our calculations have been checked by the Atlas $2$~\cite{Atlas2}, which is the differential geometry package for Mathematica). However, after we carefully analyze the full patterns of differential equations, which are completely independent of the internal coordinates, $y=(\Theta,\Phi,\psi)$, we recognize the fact that there are nontrivial choices that make the field equations solvable. Thus, the two following cases easily appear for finding exact solutions not only for the sake of simplicity but also for the deserving attention in the following subsections: (for similar choices for different manners, see, e.g., ~\cite{Mondaini,Singleton}).

\subsection{The $K(r)=K$ case}

Assume the nontrivial gauge charge $K(r)$ to be a constant, $K(r)=K$. Then, the particles still carry both electric and magnetic charge unless $J(r)=0$ or $K(r)=1$, and the particle-like solution is called a dyon solution by Schwinger~\cite{Schwinger} in Abelian theory and later by Julia$-$Zee~\cite{Julia} in non-Abelian theory by considering the 't Hooft$-$Polyakov monopole solution~\cite{Hooft1,Polyakov1}, which is the first well-known magnetic-monopole solution for the Yang$-$Mills$-$Higgs theory with gravity neglected in the literature. The dyons can also be interpreted as the excited states of magnetic monopoles from the point of view of modern physics. Then, we obtain $21$ very complicated dyon equations, and the simplest nonvanishing one becomes the following:
\begin{eqnarray}\label{K1}
 \fl l^4 \varphi(r)(K-1)\biggl\{g^2 r^2\bigl[3 J(r)^2-4 f(r) (K-1)^2\bigr] -J(r)^2 (K-1)^2 \varphi(r)^2 -\nonumber\\
\fl-J(r)^2 \bigl[g^2 r^2+(K-1)^2 \varphi(r)^2\bigr] \cos2\theta\biggr\}
+2(l^2-\ell^2) r\cos^{2}\theta (K-1) \varphi(r)J(r)J'(r)\times\nonumber\\
\fl \times\biggl\{l^2 \bigl[g^2 r^2+(K-1)^2 \varphi(r)^2\bigr]
+3 gl^2 \ell(K-1)  \varphi(r)^2+\ell^2\bigl[g^2 r^2+(K-1)^2 \varphi(r)^2\bigr] \biggr\}+\nonumber\\
\fl+\ell \varphi(r) \Biggl(g l^4 \biggl\{4 g^2 r^2\bigl[J^2-2 f(r) (K-1)^2\bigr]
-3\cos2\theta J(r)^2 (K-1)^2 \varphi(r)^2-\nonumber\\
\fl-3 J(r)^2 (K-1)^2 \varphi(r)^2  \biggr\}
+\ell(K-1)  \biggl\{-8 g^4 l^4 r^2 f(r)-g l^2 \ell(K-1) \bigl [4 g^2r^2 f(r)
-\nonumber\\
\fl-3 J(r)^2 \varphi(r)^2-3 J(r)^2 \varphi(r)^2 \cos2\theta\bigr] +2 \cos^{2}\theta \ell^2 J(r)^2 \bigl[g^2 r^2+(K-1)^2 \varphi(r)^2\bigr] \biggr\}\Biggr)+\nonumber\\
\fl+4 g^2 l^2 r^3 \varphi'(r) \Biggl( l^2 (K-1) \bigl[-J(r)^2+f(r) (K-1)^2\bigr]+ \nonumber\\
\fl+g \ell \biggl\{l^2\bigl[-J(r)^2+2 f(r) (K-1)^2\bigr]
+\ell f(r) (K-1)  \bigl[2 g l^2+\ell (K-1) \bigr]\biggr\}\Biggr)=0.
\end{eqnarray}
Here, we can prove that if $\ell=\pm l$, the equation (\ref{K1}) is reduced to a more simpler form as follows:
\begin{eqnarray}\label{K2}
 \fl(K-1\pm gl)\left[r\varphi'(r)-\varphi(r)\right]\left[f(r) (K-1) (K-1\pm 2gl)-J(r)^2\right]=0.
\end{eqnarray}

\subsubsection{Case I}

We can consider a case, where
\begin{eqnarray}\label{}
&&K-1\pm gl=0,\\
&&r\varphi'(r)-\varphi(r)\neq 0,\\
&&f(r) (K-1) (K-1\pm 2gl)-J(r)^2\neq 0,\label{nos}
\end{eqnarray}
then, this gives the following conditions for the constant $K$:
\begin{eqnarray}\label{magnetic}
K=
\cases{
1-gl & if~ $\ell=l,$\\
1+gl & if~ $\ell=-l,$\\}
\end{eqnarray}
where the both conditions, $\ell=l$ and $\ell=-l$, correspond actually to the round three-sphere due to the $\ell^{2}=(\pm l)^{2}$. We can symbolically write
\begin{eqnarray}\label{moco}
K=1-g \ell.
\end{eqnarray}
 After that, we immediately obtain from the other field equations
\begin{eqnarray}\label{electroco}
[r J'(r)-J(r)]^{2}-g^{2}\ell^{2}=0,
\end{eqnarray}
with the supplementary equation
\begin{eqnarray}\label{sco}
\varphi'(r)=0.
\end{eqnarray}
We easily achieve the following particular solutions of equations (\ref{electroco}) and (\ref{sco}):
\begin{eqnarray}\label{sbt0}
J(r)=j_{\infty}r\pm g \ell,\qquad\qquad\qquad
\varphi(r)=\varphi_{0},
\end{eqnarray}
where $j_{\infty}$ and $\varphi_{0}$ are the  corresponding integration constants of $J(r)$ and the warping function $\varphi(r)$, respectively. The parameter $j_{\infty}$ determines the scale of $J(r)$~\cite{Lugo99}, which can be explicitly written as follows:
\begin{equation}\label{}
J(r)=
\cases{
j_{\infty}r\pm g l & if~ $\ell=l,$\\
j_{\infty}r\mp g l & if~ $\ell=-l,$\\}
\end{equation}
In that respect, the remaining field equations reduce to following  set of radial differential equations:
\begin{eqnarray}\label{}
&&f'''(r)+\frac{2}{r}f''(r)-\frac{2}{r^{2}}f'(r)
+\frac{4\ell^{2}\varphi_{0}^{2}}{r^{5}}=0,\nonumber\\
&&f''(r)+\frac{2}{r}f'(r)-\frac{\ell^{2}\varphi_{0}^{2}(l^{2}+\ell^{2})}{2l^{2}r^{4}}
+\frac{1}{\ell^{2}\varphi_{0}^{2}}=0,\nonumber\\
&&f''(r)-\frac{2}{r^{2}}f(r)+\frac{2\ell^{2}}{l^{2}r^{2}}
-\frac{\ell^{2}\varphi_{0}^{2}(l^{4}+3\ell^{4})}{2l^{4}r^{4}}=0,\nonumber\\
&&f'(r)+\frac{1}{r}f(r)-\frac{1}{r}+\frac{\ell^{6}\varphi_{0}^{2}}{2l^{4}r^{3}}
+\frac{r \ell^{2}}{2l^{4}\varphi_{0}^{2}}=0.
\end{eqnarray}
Now, it is possible to simplify differential expressions above because of the $\ell^{n}=l^{n}$ only if $n$ is an even integer. Thus, we have the following equations for round $S^{3}$ internal metric:
\begin{eqnarray}\label{}
&&f'''(r)+\frac{2}{r}f''(r)-\frac{2}{r^{2}}f'(r)
+\frac{4l^{2}\varphi_{0}^{2}}{r^{5}}=0,\nonumber\\
&&f''(r)+\frac{2}{r}f'(r)-\frac{l^{2}\varphi_{0}^{2}}{r^{4}}
+\frac{1}{l^{2}\varphi_{0}^{2}}=0,\nonumber\\
&&f''(r)-\frac{2}{r^{2}}f(r)+\frac{2}{r^{2}}
-\frac{2l^{2}\varphi_{0}^{2}}{r^{4}}
=0,\nonumber\\
&&f'(r)+\frac{1}{r}f(r)-\frac{1}{r}+\frac{l^{2}\varphi_{0}^{2}}{2r^{3}}
+\frac{r}{2l^{2}\varphi_{0}^{2}}=0.
\end{eqnarray}
After integration, these differential sets admit an exact but singular solution of the following form:
\begin{eqnarray}\label{rnds}
f(r)=1-\frac{2M}{r}+\frac{l^{2}\varphi_{0}^{2}}{2r^{2}}
-\frac{r^{2}}{6l^{2}\varphi_{0}^{2}} ,
\end{eqnarray}
where $2M>0$ is the integration constant, which is identified with the total time-independent gravitational mass of the black hole. If we define the cosmological constant, $\Lambda$, as follows:
\begin{eqnarray}\label{1cosmo}
\Lambda=\frac{1}{2 l^{2}\varphi_{0}^{2}},
\end{eqnarray}
meaning that  $\Lambda>0$ and the charge-like term of the space-time, $Z$, as follows:
\begin{eqnarray}\label{cosmo}
Z^{2}=\frac{l^{2}\varphi_{0}^{2}}{2}=\frac{1}{4\Lambda},
\end{eqnarray}
then we have the Reissner$-$Nordstr\"{o}m$-$de Sitter (RNdS)-type black hole solution with a positive cosmological constant
\begin{eqnarray}\label{rnds}
f(r)=1-\frac{2M}{r}+\frac{Z^{2}}{r^{2}}
-\frac{1}{3}\Lambda r^{2}.
\end{eqnarray}
In the present situation,  the positive cosmological constant of the asymptotically de Sitter black hole is associated with a constant warping function, $\varphi_{0}$, and the radius of the three-sphere, $l$, and the de Sitter cosmological radius can be written as $a=\sqrt{6}l\varphi_{0}$ if $\Lambda=3/a^{2}$. We remark that if $l=1$, then the constant warping function can behave as a radius of three-sphere~\cite{dereli90}. The observational data suggest that the dimensionful parameter $\Lambda$ is nonzero with an accelerating universe~\cite{Riess,Perlmutter} and its value is approximately $|\Lambda|\approx~10^{-52}m^{-2}$, ~\cite{Peebles,Kagramanova}. Then, we have $l\varphi_{0}\approx~10^{26}m$, and the dimension of the charge is $[Z]=\mbox{Length}$.  It was demonstrated that from an astrophysical point of view, the supermassive black hole (SMBH) Sagittarius A$^{\ast}$ has the theoretical maximum charge order of $10^{15} C / 10^{26} C$ for a rotating$/$nonrotating case and the observational limit of charge cannot overcome $10^{8} C$ ~\cite{Za18}. Moreover, see~\cite{Juraeva21} for work on separating the impacts of the SMBH's electric and magnetic (if they exist) charges on the behavior of the magnetized particles. Now, if we convert geometrized units to SI units for the third black hole parameter charge (see, e.g.,~\cite{Riazuelo19}), we, however, obtain the length-normalized charge-like term as $Z\approx~10^{44}C$, which is inconsistent with the upper limits of those mentioned above.

We can analyze the curvature invariant, curvature scalar and roots of $f(r)$ to understand the general structure of RNdS-type black hole solutions, as usual. The Kretschmann scalar or the Lagrangian density $\hat{\mathcal{ K}}=\hat\star{(\hat{\mathcal{R}}_{AB}\wedge \hat\star{\hat{\mathcal{R}}^{AB}})}$ is calculated as follows:
\begin{eqnarray}\label{ks0}
\hat{\mathcal{ K}}=
\frac{17}{24(l\varphi_{0})^4}
+\frac{24M^{2}
}{ r^6}
-\frac{24M
(l\varphi_{0})^2 }{ r^7}
+\frac{21(l\varphi_{0})^4}{4 r^8},
\end{eqnarray}
and the Ricci scalar $\hat{\mathcal{ R}}=\hat\iota_{\hat X_{B}}\hat\iota_{\hat X_{A}}\hat{\mathcal{R}}^{AB}$ behaves as $\hat{\mathcal{R}}=7\Lambda$. Hence, the curvature invariant in equation (\ref{ks0}), as well as the metric function $f(r)$ in equation (\ref{rnds}),
exhibits a physically essential (curvature) singularity at the origin, $r=0$, and the Ricci scalar $\hat{\mathcal{ R}}$ is finite everywhere. As $r\rightarrow \infty$, the curvature invariant becomes $\hat{\mathcal{ K}}=(17/6)\Lambda^{2}$, and for $r\neq0$, it is also  finite. Conversely, the horizon function or the lapse function $\Delta(r)$ can be written as follows:
\begin{eqnarray}\label{roots1}
\Delta(r)=-\frac{1}{3}\Lambda r^{4}+r^{2}-2Mr+Z^{2}.
\end{eqnarray}
The real roots are determined by solving the quartic polynomial function $\Delta(r)\equiv0$, and they can be ordered as $r_{1}<0<r_{-}\leq r_{+}<r_{c}$ if $Z\neq0$. That is
\begin{eqnarray}\label{}
\Delta(r)=(r-r_{1})(r-r_{-})(r-r_{+})(r-r_{c}).
\end{eqnarray}
Here, $r_{-}$ is the inner (Cauchy) horizon, $r_{+}$ is the black hole event horizon, and $r_{c}$  corresponds to the outer cosmological horizon, and the negative root $r_{1}=-(r_{c}+r_{+}+r_{-})$ is nonphysical. From equation~(\ref{roots1}),  we can deduce the explicit forms of the zeros of $\Delta(r)$ in the following forms:
\begin{eqnarray}
r_{\pm}=\Omega\pm\sqrt{\Psi^{(-)}},&\qquad\qquad\qquad
r_{c,1}=-\Omega\pm\sqrt{\Psi^{(+)}},
\end{eqnarray}
where
\begin{eqnarray}
\fl\Omega=l\varphi_{0}\sqrt{1+\left[\left(\frac{3}{2}\frac{M}{l\varphi_{0}}\right)^{2}-1\right]^{1/3}},
&\quad\quad\quad\quad
\Psi^{(\pm)}=3(l\varphi_{0})^{2}\left(1\pm\frac{M}{\Omega}\right)
-\Omega^{2}.
\end{eqnarray}
If we want to obtain a space-time with horizons but without any unphysical  naked singularity, we must restrict the value of mass $M$ to certain limits by considering $l$ and $\varphi_{0}$. For instance, we suppose that $9M^{2}/4l^{2}\varphi^{2}_{0}>1$
to obtain a positive cube root term, and obviously, $\Omega>0$. This restriction is compatible with Lake's condition, $Z^{2} <9M^{2}/8$~\cite{Lake}, as expected.

We can also construct the ``extremal" or ``cold" black hole solution where two or three roots are equal to each other. For example, if the two horizons coincide $r_{-}=r_{+}=r_{h}$ at $r_{h}$, then we have $\Psi^{(-)}=0$, which means that the mass $M$ ($M<Z$) yields $M=2l\varphi_{0}/3$. Hence, $r_{h}$ becomes $r_{h}=l\varphi_{0}$. This is a charged Nariai-type black hole solution with a horizon at $r_{h}$~\cite{Bousso}. It is easy to see that the  cosmological horizon also satisfies $r_{c}=r_{h}$. In other words, three horizons (rather  than  two, as assumed) coincide at $r_{h}=r_{ucd}=l\varphi_{0}$, and we actually obtain the ultracold ($ucd$) black hole at radius $r_{ucd}$ together with $f(r_{ucd})=f'(r_{ucd})=f''(r_{ucd})=0$, $M=2r_{ucd}/3$, and  $Z^{2}=r^{2}_{ucd}/2$, which are all compatible with the results of ~\cite{Romans}. Moreover, the surface gravity of the corresponding horizon takes a form in terms of the slope of $f(r)$ at that horizon~\cite{Hawking75} given by:
\begin{eqnarray}
\kappa_{(-,+,c)}=\frac{1}{2}\left|\frac{d f(r)}{dr}\right|_{r=r_{(-,+,c)}}.
\end{eqnarray}
For instance, the surface gravity associated with the event horizon $\kappa_{+}$ can be written as follows:
\begin{eqnarray}
\kappa_{+}=\frac{1}{2}\left|\frac{2 M}{r_{+}^2}-\frac{r_{+}}{3 (l \varphi_{0} )^2}-\frac{(l \varphi_{0} )^2}{r_{+}^3}\right|.
\end{eqnarray}
Furthermore, the associated Hawking temperature, $\mathcal{T}_{H}$,~\cite{Hawking75} can be calculated by using the relationship $\mathcal{T}_{H}=|\kappa|/2\pi$, i.e.,
\begin{eqnarray}
\mathcal{T}_{+}=\frac{1}{4\pi}\left|\frac{2 M}{r_{+}^2}-\frac{r_{+}}{3 (l \varphi_{0} )^2}-\frac{(l \varphi_{0} )^2}{r_{+}^3}\right|.
\end{eqnarray}
In the extreme case, the RNdS black hole has zero temperature, and for the $Z^{2}=M^{2}$ case, which represents lukewarm black holes~\cite{Ross95} the temperature is given by
\begin{eqnarray}
\mathcal{T}=\frac{1}{6\pi l \varphi_{0}}\sqrt{\sqrt{3}-\frac{3}{2}}\approx\frac{0.026}{l \varphi_{0}},
\end{eqnarray}
and the solutions are stable in this fashion. This temperature is also directly obtained from the relationship
\begin{eqnarray}
\mathcal{T}=\frac{1}{2\pi}\sqrt{\frac{ \Lambda}{3}\left(1-4M\sqrt{\frac{ \Lambda}{3}}\right)},
\end{eqnarray}
which is given in ~\cite{Romans}.

We can also record the gauge potentials in the following forms:
\begin{eqnarray}\label{gauge1}
A^{\alpha}=\left(\pm \frac{q}{r^{2}}+\frac{j_{\infty}}{g r}\right){x}^{\alpha}dt+\frac{p}{r^{2}}\varepsilon^{\alpha}\,_{ab}{x}^{b}dx^{a}.
\end{eqnarray}
Here, the constant electric charge-like term  $q=\ell$  and the constant magnetic charge-like term $p=-\ell$ are placed into the general gauge equation above for later convenience. In fact, because of the quantization conditions, we can already write $p=(1/2)nR$, where $R$ is the KK circle radius of $S^{1}$ for the 5D Abelian model~\cite{Angus1}. Thus, it is not surprising that the magnetic monopole charge $p$ is related to the internal space parameter $\ell$. In the first place, it is not difficult to conjecture that $A_{(\ell=-l)}^{\alpha}\neq-A_{(\ell=l)}^{\alpha}$. However, by considering Dirac's quantization relation~\cite{dirac}, we require not only that $e=g$ but also that $g=-(1/\ell)$, which satisfies ${\mathcal{D}}\star{F^{\alpha}}=0$. Then, the gauge potentials which satisfies $A_{(\ell=-l)}^{\alpha}=-A_{(\ell=l)}^{\alpha}$ can be explicitly written as follows:
\begin{eqnarray}\label{pote}
\left( \begin{array}{c}
   A^{1}   \\
         A^{2} \\
        A^{3} \\
   \end{array} \right)&=&\mathcal{J}(r)
\left( \begin{array}{c}
   \sin\theta\cos\phi   \\
         \sin\theta\sin\phi \\
        \cos\theta \\
   \end{array} \right)dt+\nonumber\\
   &&+\mathcal{K}(r)\left[\left( \begin{array}{c}
   \sin\phi   \\
         -\cos\phi \\
       0 \\
   \end{array} \right)d\theta
   -\sin\theta
   \left( \begin{array}{c}
   \cos\theta\cos\phi  \\
        \cos\theta\sin\phi \\
       \sin\theta \\
   \end{array} \right)d\phi\right],
\end{eqnarray}
where
\begin{eqnarray}\label{}
\mathcal{J}(r)=\frac{{J}(r)}{gr}=\pm \frac{ q}{r}-\ell j_{\infty},   &  \qquad\qquad\qquad \mathcal{K}(r)=\frac{{K}(r)-1}{g}=p.
\end{eqnarray}
Moreover, when $J(r)=0$, the gauge potentials reduce to the well-known Wu$-$Yang monopole with infinite classical energy~\cite{wuyang,stuller}, but this case is not a solution of our model.

For $SU(2)$ gauge fields, the dimensionless, global, gauge invariant non-Abelian electric charge,  ${\mathcal{Q}}^{{YM}}$, and the magnetic charge, ${\mathcal{P}}^{{YM}}$, may be defined by applying surface integrals over the two-sphere $S^{2}$ at $r\rightarrow\infty$, as in ~\cite{Ashtekar1,Ashtekar2,Corichi1,Corichi2}:
\begin{eqnarray}\label{chargen}
\fl{\mathcal{Q}}^{{YM}}\equiv-\frac{g}{4\pi}\oint_{r\rightarrow\infty}\sqrt{\sum_{\alpha}
(\star F^{\alpha}_{\theta\phi})^{2}} d\theta d\phi,
\quad\quad
{\mathcal{P}}^{{YM}}\equiv-\frac{g}{4\pi}\oint_{r\rightarrow\infty}\sqrt{\sum_{\alpha}
(F^{\alpha}_{\theta\phi})^{2}} d\theta d\phi.
\end{eqnarray}
The equations above are also useful for obtaining horizon charges and local charges~\cite{Kleihaus1,Kleihaus2}. Hence, following the YM topological charges from (\ref{chargen}), we have
\begin{eqnarray}\label{}
 {\mathcal{Q}}^{{YM}}=-g|q|,  &   \qquad\qquad\qquad   {\mathcal{P}}^{{YM}}=-g|p|,\label{mc}
\end{eqnarray}
where $|q|=|p|=l$, which confirm that $q$ and $p$  are related to the ${\mathcal{Q}}^{{YM}}$ and ${\mathcal{P}}^{{YM}}$ of the space-time, respectively. That is
\begin{eqnarray}\label{dyoncharge}
\fl\mathcal{Q}^{{YM}}=
\cases{
+1 & if~ $\ell=l,$\\
-1 & if~ $\ell=-l,$\\
}
&\qquad\qquad\qquad
\mathcal{P}^{{YM}}=
\cases{
+1 & if~ $\ell=l,$\\
-1 & if~ $\ell=-l.$\\
}
\end{eqnarray}
Thus, it is found that for the case $\ell=l$ the gauge field corresponds to the positively charged dyon solution, whereas  we have the negatively charged antidyon solution for the case $\ell=-l$. It seems that whether a dyon or antidyon solution is obtained depends on the sign of the internal space-squashing parameter, $\ell$. In this fashion, the (anti-)dyonic black hole carries one unit of electric charge and one unit of magnetic charge, and it is appropriate to redefine parameter $Z^{2}$ in (\ref{cosmo}) as follows:
\begin{eqnarray}\label{}
Z^{2}=\frac{{\mathcal{Q}}^{2}+{\mathcal{P}}^{2}}{8\Lambda},
\end{eqnarray}
where the total (electric plus magnetic) time-independent gravitational charge of the space-time is equal to
\begin{eqnarray}\label{charges}
{\mathcal{Q}}^{2}+{\mathcal{P}}^{2}=2.
\end{eqnarray}
Hence, we have the following equation, which is nothing but (\ref{cosmo})
\begin{eqnarray}\label{}
Z^{2}=\frac{1}{4\Lambda}.
\end{eqnarray}
It is essential here to note that for the $SU({\mathcal{N}})$ gauge field, the only value of the magnetic charge ${\mathcal{P}}$ appearing in the black hole solutions is given by (see, e.g., ~\cite{Winstanley07,Winstanley08,Winstanley12})
\begin{eqnarray}\label{}
{\mathcal{P}}^{2}=\frac{1}{6}{\mathcal{N}}({\mathcal{N}}+1)({\mathcal{N}}-1).
\end{eqnarray}
Hence, for the gauge group ${\mathcal{G}}=SU(2)$, we have ${\mathcal{P}}^{2}=1$, which we already obtained from equation (\ref{dyoncharge}). We notice that if $g\neq -(1/\ell)$, the total charge becomes ${{\mathcal{Q}}}^{2}+{{\mathcal{P}}}^{2}=2g^{2}q^{2}$. Then, the new modified cosmological constant $\widetilde{\Lambda}$ and the charge term $\widetilde{Z}^{2}$ become in the following forms:
\begin{eqnarray}\label{}
\widetilde{\Lambda}=\frac{g^{2}}{2\varphi_{0}^{2}}=g^{2}q^{2}\Lambda,
\qquad\qquad\qquad
\widetilde{Z}^{2}=\frac{{{\mathcal{Q}}}^{2}
+{{\mathcal{P}}}^{2}}{8\widetilde{\Lambda}}
=\frac{g^{2}q^{2}}{4\widetilde{\Lambda}}.
\end{eqnarray}
The new positive cosmological term $\widetilde{\Lambda}$ depends on the constant warping function, $\varphi_{0}$, as well as the charge parameter, $g$, in this case. It is not difficult to conjecture that these equations reduce to the previous equations (\ref{1cosmo}) and (\ref{cosmo}), if and only if $g=\pm(1/\ell)$, respectively.

It is useful a quickly review  the main results of the remarkable studies of Perry~\cite{Perry} and Angus~\cite{Angus}. They considered a case where not only was the field strength in a four-index form, $\hat F_{ABCD}$, but also the gauge potentials were simpler than the generalized Wu$-$Yang ansatz (\ref{WuYangansatz}), by considering the 6D and 7D Einstein equations together with a cosmological constant and the matter fields, respectively. In the compact form, the gauge ansatz that was used was $A^{\alpha}=\delta^{\alpha}_{3}A(r,\theta)$, which can be explicitly written as follows:
\begin{eqnarray}\label{potee}
\left( \begin{array}{c}
   A^{1}   \\
         A^{2} \\
        A^{3} \\
   \end{array} \right)=
\mathcal{J}(r)\left( \begin{array}{c}
   0  \\
         0 \\
         -1\\
   \end{array} \right)dt
   +\mathcal{K}(r) \left( \begin{array}{c}
   0  \\
        0 \\
       \cos\theta  \\
   \end{array} \right)d\phi.
\end{eqnarray}
This condition is also known as the Abelian gauge, because of its similarity to the $U(1)$  Abelian model~\cite{Marciano}, keeping in mind that any Abelian solutions solve the non-Abelian equations. Perry found the Abelian-like solution to the 6D model with a matter source from the Einstein action in the following form:
\begin{eqnarray}\label{rnds1}
\fl f(r)=1-\frac{2M}{r}+\frac{Z^{2}}{r^{2}},\qquad\qquad
\Lambda=\frac{1}{2\varphi_{0}^{2}},\qquad\qquad
Z^{2}=\frac{q^{2}+p^{2}}{8\Lambda}=\frac{p^{2}}{4\Lambda},
\end{eqnarray}
with $q=p$ and $p$ is a constant.
The gauge field and whose functions are
\begin{eqnarray}\label{gfperry}
\fl A^{3}=\mp \frac{q}{r}dt+p\cos\theta d\phi,\qquad \qquad\qquad
\mathcal{J}(r)=\pm \frac{q}{r}, \qquad\qquad\qquad    \mathcal{K}(r)=p.
\end{eqnarray}
The topological charge, $p$, must be $p=(n/2)$, where $n$ is an integer (for $Z_{2}$ monopoles $p=\pm (1/2)$,~\cite{Preskill}) to avoid string singularities. Conversely, the solutions of the 7D model found by Angus are very problematic, and only approximate solutions may be possible, because of the more complicated non-Abelian KK metric ansatz. The metric function, $f(r)$, can roughly be written as follows:
\begin{eqnarray}\label{rnds2}
\fl f(r)=1-\frac{2M}{r}+\frac{Z^{2}}{r^{2}}+O(\frac{1}{r^{3}}),
\qquad\qquad
\Lambda=\frac{5}{4l^{2}},\qquad\qquad
Z^{2}=\frac{q^{2}+p^{2}}{4\Lambda},
\end{eqnarray}
for $\alpha_{0}=0$. Furthermore, the gauge field  and the unknown profile functions respectively are
\begin{eqnarray}\label{gfangus}
\fl A^{3}=-\left[\frac{q}{r}+O(\frac{1}{r^{2}})\right]dt+p\cos\theta d\phi, \qquad\quad
\mathcal{J}(r)=\frac{q}{r}+O(\frac{1}{r^{2}}),   \qquad\quad  \mathcal{K}(r)=p,
\end{eqnarray}
where $q$ and $p$ are again constant.

The nontrivial solution (\ref{rnds}), which is new for the considered model in the literature, turns out to be a 4D vacuum solution of Einsteinian gravity (see, e.g.,~\cite{Cotaescu}), i.e., ${\hat{\mathcal{R}}}_{A}=\hat\Lambda\hat E_{A}$ for all constant $\hat\Lambda$ where $\hat\Lambda=\Lambda$, which corresponds to the Einstein manifold, if we start with assuming the ansatzes (\ref{moco}) and (\ref{sbt0}). The cosmological constant comes naturally out from the field equations as an integration constant in the WYKK theory. ``This situation seems to be more natural than adding a cosmological constant into the Einstein field equations by hand.'' according to Chen et al. ~\cite{chen1} (see also ~\cite{cook}).  Moreover, not only is the solution (\ref{rnds})  in a more desirable and suitable form than  both Perry's (\ref{rnds1}) and Angus' (\ref{rnds2}) solutions, but also the  spherically symmetric gauge potentials (\ref{pote}) are more general and non-Abelian than both (\ref{gfperry}) and (\ref{gfangus}) (see Table~\ref{ta1}).
\begin{table}
\centering
\caption{\label{ta1} Comparison of the functions of three theory models.}
\lineup
\begin{tabular}{@{}*{3}{l}}
\br
WYKK~in~7D&PERRY~in~6D&ANGUS~in~7D\cr
\mr
$\hat D_{A}\hat R^{A}\,_{BCD}=0$&$\hat G_{AB}+\hat\Lambda\hat g_{AB}=\hat T_{AB}[\hat F]^{\rm a}$
&$\hat G_{AB}+\hat\Lambda\hat g_{AB}=\hat T_{AB}[\hat F]^{\rm a}$\cr
$f(r)=1-\frac{2M}{r}+\frac{Z^{2}}{r^{2}}-\frac{1}{3}\Lambda r^{2}$
& $f(r)=1-\frac{2M}{r}+\frac{Z^{2}}{r^{2}}$
&$f(r)=1-\frac{2M}{r}+\frac{Z^{2}}{r^{2}}+O(\frac{1}{r^{3}})$\cr
$\Lambda=\frac{1}{2 l^{2}\varphi_{0}^{2}}$
& $\Lambda=\frac{1}{2\varphi_{0}^{2}}$
&$\Lambda=\frac{5}{4l^{2}}$\cr
$Z^{2}=\frac{1}{4\Lambda}$
&$Z^{2}=\frac{p^{2}}{4\Lambda}$
&$Z^{2}=\frac{q^{2}+p^{2}}{4\Lambda}$\cr
$\varphi(r)=\varphi_{0}$
&$\varphi(r)=\varphi_{0}$
&$\varphi(r)=1+O(\frac{1}{r^{4}})$\cr
$\begin{array}{r@{}l@{}}
A^{\alpha}& {}=\frac{\mathcal{J}(r)}{r}{x}^{\alpha}dt+\\
& {}+\frac{\mathcal{K}(r)}{r^{2}}\varepsilon^{\alpha}\,_{ab}{x}^{b}dx^{a}\end{array}$
&$\begin{array}{r@{}l@{}}
A^{1}& {}=A^{2}=0 \\ A^{3}& {}=-\mathcal{J}(r)dt+\\
& {}+\mathcal{K}(r) \cos\theta d\phi\end{array}$
&$\begin{array}{r@{}l@{}}
A^{1}& {}=A^{2}=0 \\ A^{3}& {}=-\mathcal{J}(r)dt+\\
& {}\mathcal{K}(r) \cos\theta d\phi\end{array}$\cr
$\mathcal{J}(r)=\pm \frac{ q}{r}-\ell j_{\infty}$
&$\mathcal{J}(r)=\pm \frac{q}{r}$
&$\mathcal{J}(r)=\frac{q}{r}+O(\frac{1}{r^{2}})$\cr
$\mathcal{K}(r)=p$
&$\mathcal{K}(r)=p$
&$\mathcal{K}(r)=p$\cr
$q=-p=\ell$
&$q=p$
&$q$~ and~ $p$~ are~ any~ constant\cr
\br
\end{tabular}
 $^{\rm a}\hat T_{AB}[\hat F]=\hat F_{ACDE}\hat F_{B}\,^{CDE}-(1/8)\hat g_{AB}\hat F^{2}$.
\end{table}
Furthermore, we have the components of the gravitational energy-momentum second-rank tensor, which are given by applying the RNdS-like black hole solution and expressing the relationship $\hat T_{AB}=\hat\iota_{\hat X_{B}}\hat\star{\hat \tau_{\hat X_{A}}}$ in the following forms:
\begin{eqnarray}\label{angle3}
\fl\hat T_{tt}=\frac{3}{16(l\varphi_{0})^4}+\frac{2r^4+24(l\varphi_{0})^2 (3r^{2}-8 M r)+63 (l\varphi_{0})^4}{24 r^8 },\nonumber\\
\fl\hat T_{rr}=-\frac{3}{16(l\varphi_{0})^4}-\frac{10 r^4 +24 (l\varphi_{0})^2( r^2 -4  Mr)
+39 (l\varphi_{0})^4}{24 r^8},\nonumber\\
\fl\hat T_{\theta\theta}=\hat T_{\phi\phi}=-\frac{3}{16(l\varphi_{0})^4}+\frac{2r^4
+16 (l\varphi_{0})^2( r^2-3 Mr)+15 (l\varphi_{0})^4}{8 r^8},\nonumber\\
\fl\hat T_{\Theta\Theta}=-\frac{5}{48(l\varphi_{0})^4}-\frac{96M \left[Mr^{2}-(l\varphi_{0})^2 r\right]+(l\varphi_{0})^4 \left(21-4\sin^{2}\theta\cos^{2}\phi\right)}{8 r^8 },\nonumber\\
\fl\hat T_{\Phi\Phi}=-\frac{5}{48(l\varphi_{0})^4}-\frac{96M \left[Mr^{2}-(l\varphi_{0})^2 r\right]+(l\varphi_{0})^4 \left(21-4\sin^{2}\theta\sin^{2}\phi\right)}{8 r^8 },\nonumber\\
\fl\hat T_{\psi\psi}=-\frac{5}{48(l\varphi_{0})^4}-\frac{96M \left[Mr^{2}-(l\varphi_{0})^2 r\right]+(l\varphi_{0})^4(19-2 \cos2\theta)}{8 r^8}.
\end{eqnarray}
Conversely, the flux-like components can be found to be
\begin{eqnarray}\label{}
&\frac{\hat T_{t\Theta}}{\sin\theta\cos\phi}=\frac{\hat T_{t\Phi}}{\sin\theta\sin\phi}=\frac{\hat T_{t\psi}}{\cos\theta}
=\frac{\ell \varphi_{0} \sqrt{f(r)} \left[6Mr-5 (l \varphi_{0})^2\right]}{r^7},\label{timet}\\
&\frac{\hat T_{\Theta\Phi}}{\sin^{2}\theta\sin2\phi}=\frac{\hat T_{\Theta\psi}}{\sin2\theta\cos\phi}=\frac{\hat T_{\Phi\psi}}{\sin2\theta\sin\phi}=
\frac{(l\varphi_{0})^{4}}{4r^{8}}\label{angle1}.
\end{eqnarray}
One can also write the remaining components of the stress-energy tensor by applying its symmetry  properties. All other components vanish, as well. Let us finally mention that the all stress-energy tensors above are divergence at the singular point $r=0$. The components in equation (\ref{timet}) that depend only on the parameter $\ell$, among all the components, vanish at the zeros of $f(r)$. Thus, to make all off-diagonal components $0$, we consider the following proper values for $\theta$ and $\phi$ in (\ref{angle1})
\begin{eqnarray}
\fl S_{(\theta,\phi)}=\left\{
(0,\phi),~ (\frac{\pi}{2},0),~
(\frac{\pi}{2},\frac{\pi}{2}),~ (\frac{\pi}{2},\pi),~ (\frac{\pi}{2},\frac{3\pi}{2}),~ (\frac{\pi}{2},2\pi),~ (\pi,\phi)\right\},
\end{eqnarray}
and $r=r_{+}$ in (\ref{timet}). Now, all off-diagonal components vanish.
Meaning also that the $\hat T_{\Theta\Theta}$, $\hat T_{\Phi\Phi}$ and $\hat T_{\psi\psi}$ depend only on the $r$. Then, the gravitational energy-momentum tensor, $\hat T_{AB}$, is just diagonal and is formally similar to that of Maxwell's theory of electromagnetism.

\subsubsection{Case II}

We already obtain the following condition from the equation (\ref{K2}), which do not appear in the empty space solutions of the Einstein field equation:
\begin{eqnarray}\label{dilatone}
\varphi'(r)-\frac{1}{r}\varphi(r)=0.
\end{eqnarray}
The solution of warping function (\ref{dilatone}) is found to be
\begin{eqnarray}\label{sbt2}
\varphi(r)=\varphi_{0}r,
\end{eqnarray}
where $\varphi_{0}$ is any constant. It is straightforward  to insert solution (\ref{sbt2}) into the field equations to investigate the exact analytic solutions. Doing this, one of the field equations becomes the following:
\begin{eqnarray}\label{}
\fl rJ^{2}(r)\frac{f'(r)}{f(r)}-(K-1)(K-1+2g\ell)f(r)-2rJ(r)J'(r)
+J^{2}(r)=0.
\end{eqnarray}
It may be rewritten as follows:
\begin{eqnarray}\label{}
\left[\frac{J^{2}(r)}{r f(r)}\right]'-\frac{(K-1)(K-1+2g\ell)}{r^{2}}=0.
\end{eqnarray}
Hence, we have
\begin{eqnarray}\label{nonE}
f(r)=\frac{J^{2}(r)}{f_{0} r+(K-1)(K-1+2g\ell)},
\end{eqnarray}
where $r\neq-(K-1)(K-1+2g\ell)/f_{0}$ and $f_{0}$ is an integration constant, $f_{0}\neq 0$. The solution (\ref{nonE}) can reduce to the solution (\ref{nos}) if $f_{0}=0$, i.e.,
\begin{eqnarray}\label{K3}
f(r)=\frac{J^{2}(r)}{(K-1)(K-1+2g\ell)},
\end{eqnarray}
where $K\neq 1$ and $K\neq 1-2g\ell$. We may classify equations (\ref{nonE}) and (\ref{K3}) as physical or unphysical  non-Einsteinian solutions. In both cases, however, we fail to obtain exact solutions, which satisfies all of the remaining coupled field equations, even if we consider further assumptions regarding the functions $J(r)$ and $K(r)=K$ .

\subsection{The $J(r)=0$ case}

We can alternatively consider the case where the Coulomb part of the gauge field is chosen so that ${A}^{\alpha}_{0}(\vec{x})=0$. In other words, the electric field vanishes at $J(r)\equiv0$, and the YM fields become purely magnetic, rather than dyons, because of this restricted monopole configuration. In this  regard, the corresponding magnetic-monopole solutions are dipole-like solutions, i.e., $F^{\alpha}_{ab}\sim r^{-3}$ (see the equation~(\ref{YMF})), and they  vanish faster than the monopole-like solutions, i.e., $F^{\alpha}_{ab}\sim r^{-2}$. Then, the simplest field equation becomes
\begin{eqnarray}\label{J1}
\fl 3 g^2 l^2 \ell^2 r^2 f(r)\varphi(r)^2 [K'(r)]^2
+2 g^2 l^2 (l^2-\ell^2) r^2 f(r)\varphi(r)\varphi'(r) [K(r)-1]K'(r)-\nonumber\\
 \fl-3 [K(r)-1]^2 \Biggl(-g^2 l^2 r^2+2 g l^2 \ell  [K(r)-1] \varphi(r)^2+\ell^{2} \biggl\{[K(r)-1]^2 \varphi(r)^2+g^2r^2
+\nonumber\\
 \fl+4g^2 l^2 \varphi(r)^2
+2 g \ell[K(r)-1] \varphi(r)^2 \biggr\}\Biggr)=0.
\end{eqnarray}
Again, we can assume that $\ell=\pm l$, then the equation (\ref{J1}) simply gives
\begin{eqnarray}\label{J2}
 \varphi(r)^2 \left\{g^2 l^2 r^2 f(r) \left[K'(r)\right]{}^2-[K(r)-1]^2 [K(r)-1\pm2gl]^2\right\}=0.
\end{eqnarray}
Thus, the metric function $f(r)$ takes the form if $\varphi(r)\neq 0$ as follows:
\begin{eqnarray}\label{none3}
 f(r) =\frac{[K(r)-1]^2 [K(r)-1\pm2gl]^2}{g^2 l^2 r^2\left[K'(r)\right]{}^2}.
\end{eqnarray}
However, we could not find any solution, which satisfies all of the remaining coupled field equations for $K(r)\neq constant$. Hence, let us assume again that the space component function of the gauge field is of the form $K(r)=K$, where $K$ is a constant and $K\neq 1$, to simplify the field equations. Then, using $\ell^{n}=l^{n}$ with $n$ being an even integer, we obtain by applying the equation (\ref{J2})
\begin{eqnarray}\label{monopole}
K=1- 2g \ell.
\end{eqnarray}
Here, we also prove that $K=1- g \ell$ is surprisingly not a solution anymore for this considered case. The piecewise-defined constant $K$ can explicitly be written in the following forms:
\begin{equation}\label{magnetic}
K=
\cases{
1-2gl & if~$\ell=l,$\\
1+2gl & if~$\ell=-l,$\\
}
\end{equation}
for the round internal metric. The gauge potentials, which are electrically neutral are found to be
\begin{eqnarray}\label{}
\left( \begin{array}{c}
   A^{1}   \\
         A^{2} \\
        A^{3} \\
   \end{array} \right)=\mathfrak{p}\Bigl[
   \left( \begin{array}{c}
   \sin\phi   \\
         -\cos\phi \\
       0 \\
   \end{array} \right)d\theta
   +\sin\theta
   \left( \begin{array}{c}
   \cos\theta\cos\phi  \\
        \cos\theta\sin\phi \\
       -\sin\theta \\
   \end{array} \right)d\phi\Bigr],
\end{eqnarray}
which again satisfy  $A_{(\ell=-l)}^{\alpha}=-A_{(\ell=l)}^{\alpha}$ with the constant magnetic charge-like term $\mathfrak{p}=-2\ell$. In the round limit, all components of the YM strength (\ref{YMF}) vanish, and the $A_{(\ell=l)}^{\alpha}$ is the so-called pure gauge. However, there are two possibilities for the constant $g$  that satisfy the 4D source-free YM equations ${\mathcal{D}}\star F^{\alpha}=0$ with $e=g$ for this situation. These are $g=-(1/\ell)$ and $g=-(1/2\ell)$. The electric charge already vanishes, ${\mathcal{Q}}^{{YM}}=0$, in the absence of $J(r)$, whereas ${\mathcal{P}}^{YM}$ gives zero charge in both the $\ell=l$ and $\ell=-l$ cases,  for $g=-(1/\ell)$. Conversely, for $g=-(1/2\ell)$, the nonzero color magnetic charge ${\mathcal{P}}^{{YM}}$ (\ref{chargen}) is calculated as follows:
\begin{eqnarray}\label{}
{\mathcal{P}}^{{YM}}=-g|\mathfrak{p}|.
\end{eqnarray}
Then, we obtain the positively charged monopole and the negatively charged antimonopole as follows:
\begin{equation}\label{dyoncharge2}
 \mathcal{P}^{{YM}}=
 \cases{
 +1 & if~ $\ell=l,$\\
-1 & if~ $\ell=-l.$\\
}
\end{equation}
This monopole solution, which has zero energy, was previously discussed by Wu and Yang by considering the sourceless YM equations. However, the antimonopole solution is new and corresponds to the infinite energy Wu$-$Yang monopole~\cite{wuyang}.

Next, we  turn to the investigation of the black hole solutions. Inserting expression~(\ref{monopole}) into the field equations yields a system of differential equations comprising a homogeneous linear equation of the third order and two nonhomogeneous linear equations of the second order with variable coefficients as follows:
\begin{eqnarray}\label{}
\fl f'''+\left[3\left(\frac{\varphi'}{\varphi}\right)+\frac{2}{r}\right]f''
-\left[3\left(\frac{\varphi'}
{\varphi}\right)^{2}+\frac{2}{r^{2}}\right]f'=0,\label{monopolefield1}\\
\fl f''+3\left(\frac{\varphi'}{\varphi}\right)f'
-2\left[3\left(\frac{\varphi'}
{\varphi}\right)^{2}+\frac{1}{r^{2}}\right]f+\frac{2\ell^{2}}{l^{2}r^{2}}
=0,\label{monopolefield2}\\
\fl\left(\frac{\varphi'}{\varphi}\right)\Bigg\{f''
+\Biggl[
4\left(\frac{\varphi}{\varphi'}\right)\left(\frac{\varphi'}{\varphi}\right)'
+5\left(\frac{\varphi'}{\varphi}\right)+\frac{2}{r}
\Biggr]f'+\nonumber\\\fl
+2\Biggl\{
\left(\frac{\varphi}{\varphi'}\right)\left(\frac{\varphi'}{\varphi}\right)''
+4\left(\frac{\varphi'}{\varphi}\right)'
+\frac{2}{r}\Biggl[\left(\frac{\varphi}{\varphi'}\right)\left(\frac{\varphi'}{\varphi}\right)'
+\left(\frac{\varphi'}{\varphi}\right)-\frac{1}{r}\Biggr]
\Biggr\}f
+\frac{1}{l^{2}\varphi^{2}}\Bigg\}=0.\label{monopolefield3}
\end{eqnarray}
We recognize the fact that there is no relationship between the charge-like constant $\mathfrak{p}$ and the field equations above. Hence, this case may be considered as a kind of Brans$-$Dicke-type case. In the first place, it is not easy to obtain the general analytic solutions of two unknown functions $f=f(r)$ and $\varphi=\varphi(r)$ from the field equations~(\ref{monopolefield1})--(\ref{monopolefield3}). However, the order of equation (\ref{monopolefield1}) can be reduced by the substitution $f'(r)=\mathbb{f}(r)$,
\begin{eqnarray}\label{frobenius}
 \mathbb{f}''(r)+\mathbb{p}(r)\mathbb{f}'(r)+\mathbb{q}(r)\mathbb{f}(r)=0,
\end{eqnarray}
 where
\begin{eqnarray}\label{ks}
\mathbb{p}(r)=3\left(\frac{\varphi'}{\varphi}\right)+\frac{2}{r},\qquad \qquad \qquad \mathbb{q}(r)=-3\left(\frac{\varphi'}
{\varphi}\right)^{2}-\frac{2}{r^{2}},
\end{eqnarray}
which means that, there is at least one nontrivial solution to the equation (\ref{frobenius}), of the form:
\begin{eqnarray}\label{fere}
 \mathbb{f}(r)=r^{\sigma}\sum^{\infty}_{n=0}\mathbb{f}_{n}r^{n}
 =\mathbb{f}_{0}r^{\sigma}+\mathbb{f}_{1}r^{\sigma+1}+\mathbb{f}_{2}r^{\sigma+2}+...
\end{eqnarray}
It is called  the well-known Frobenius power series~\cite{Arfken1985}, where the exponent $\sigma$ may be a real or complex number, which must be determined with $\mathbb{f}_{n}$ obtained from the recursive relation and $\mathbb{f}_{0}\neq0$. Clearly, $r=0$ be a regular singular point of the equation (\ref{frobenius}), then $r\mathbb{p}(r)$ and $r^{2}\mathbb{q}(r)$ are analytics at $r=0$, so that we may write
\begin{eqnarray}\label{f11}
r\mathbb{p}(r)&=&\sum^{\infty}_{n=0}\mathbb{p}_{n}r^{n}=\mathbb{p}_{0}+\mathbb{p}_{1}r
+\mathbb{p}_{2}r^{2}+...,\\
r^{2}\mathbb{q}(r)&=&\sum^{\infty}_{n=0}\mathbb{q}_{n}r^{n}=\mathbb{q}_{0}+\mathbb{q}_{1}\label{f12}
+\mathbb{q}_{2}r^{2}+....
\end{eqnarray}
The $\mathbb{q}_{n}$ terms can be calculated in terms of parameters $\mathbb{p}_{n}$ order by order. Then, the first three terms can be written as follows:
\begin{eqnarray}\label{kus}
\fl\mathbb{q}_{0}=-\frac{\mathbb{p}^{2}_{0}-4\mathbb{p}_{0}+10}{3},\qquad
\mathbb{q}_{1}=-\frac{2(\mathbb{p}_{0}-2)\mathbb{p}_{1}}{3},\qquad
\mathbb{q}_{2}=-\frac{2(\mathbb{p}_{0}-2)\mathbb{p}_{2}+\mathbb{p}^{2}_{1}}{3}.
\end{eqnarray}
Using the equation (\ref{ks}), we have
\begin{eqnarray}\label{fibölüfi}
\frac{\varphi'(r)}{\varphi(r)}=\frac{1}{3}\left[\frac{\mathbb{p}_{0}-2}{r}+\mathbb{p}_{1}
+\mathbb{p}_{2}r+\mathbb{p}_{3}r^{2}+...
\right],
\end{eqnarray}
that is also the coefficient function of the last field equation (\ref{monopolefield3}). Then, we can obtain that the warping function is a series of the form:
\begin{eqnarray}\label{scalarserie}
\fl\varphi(r)&=&\varphi_{0}\left[r^{\frac{\mathbb{p}_{0}-2}{3}}+\frac{\mathbb{p}_{1}}{3}r^{\frac{\mathbb{p}_{0}+1}{3}}
+\frac{(\mathbb{p}^{2}_{1}+3\mathbb{p}_{2})}{18}r^{\frac{\mathbb{p}_{0}+4}{3}}
+\frac{(\mathbb{p}^{3}_{1}+9\mathbb{p}_{1}\mathbb{p}_{2}
+18\mathbb{p}_{3})}{162}r^{\frac{\mathbb{p}_{0}+7}{3}}+...\right],
\end{eqnarray}
where $\varphi_{0}$ is any constant. Conversely, if the series  $\mathbb{f}(r)$  (\ref{fere}), $\mathbb{p}(r)$ (\ref{f11}) and $\mathbb{q}(r)$  (\ref{f12}) are inserted into the field equations~(\ref{monopolefield1}) and (\ref{monopolefield2}), where
\begin{eqnarray}\label{intfr}
f(r)=f_{0}+\int\mathbb{f}(r)dr,
\end{eqnarray}
and $f_{0}$ is a constant of integration, then we have two indicial equations, $F_{1}$ and $F_{2}$, with two unknowns, $\sigma$ and $\mathbb{p}_{0}$, from the coefficients of the lowest power of $r$, as follows:
\begin{eqnarray}\label{s1}
F_{1}(\sigma,\mathbb{p}_{0})&=&3\sigma^{2}+3(\mathbb{p}_{0}-1)\sigma-\mathbb{p}_0^2+4 \mathbb{p}_0-10=0,\nonumber\\
F_{2}(\sigma,\mathbb{p}_{0})&=&3 \sigma ^2+3\left(\mathbb{p}_0-1\right) \sigma -2 \mathbb{p}_0^2+11 \mathbb{p}_0-20=0,
\end{eqnarray}
where $\sigma\neq-1$. We can successively find the solution set, $S$, of the system~(\ref{s1}) as follows:
\begin{eqnarray}
\fl S=\{
(\sigma \to -2,\mathbb{p}_0\to 2),~ (\sigma \to 1,\mathbb{p}_0\to 2),~
(\sigma \to -5,\mathbb{p}_0\to 5),~ (\sigma \to 1,\mathbb{p}_0\to 5)\}.
\end{eqnarray}
Note that for both $\mathbb{p}_0= 2$ and $\mathbb{p}_0= 5$, the roots differ by a positive integer, $\sigma_{1}-\sigma_{2}=N$, then there always exist two linearly independent
solutions. Hence, we consider each of these cases in detail in the following subsections:

\subsubsection{The $\sigma_{2} =-2$ and $\mathbb{p}_0=2$ case}

First, we restrict ourselves to the case, where the smaller root is $\sigma_{2} =-2$ and $\mathbb{p}_0=2$. Then, by applying the equations~(\ref{fere}) and~(\ref{scalarserie}), we have
\begin{eqnarray}\label{sigmaeksiiki1}
\fl\mathbb{f}(r)
=\frac{\mathbb{f}_{0}}{r^{2}}+\frac{\mathbb{f}_{1}}{r}+\mathbb{f}_{2}+\mathbb{f}_{3}r+\mathbb{f}_{4}r^{2}+...,\\
\fl\varphi(r)=\varphi_{0}\left[1+\frac{\mathbb{p}_{1}}{3}r
+\frac{(\mathbb{p}^{2}_{1}+3\mathbb{p}_{2})}{18}r^{2}
+\frac{(\mathbb{p}^{3}_{1}+9\mathbb{p}_{1}\mathbb{p}_{2}+18\mathbb{p}_{3})}{162}r^{3}+...\right].\label{sigmaeksiiki2}
\end{eqnarray}
It is straightforward to verify from equations~(\ref{monopolefield1}) and (\ref{monopolefield2}) that the recursion equation leads to
\begin{eqnarray}
\fl\mathbb{f}_{0}=\mathbb{f}_{0},\qquad \qquad\mathbb{f}_1=-\mathbb{f}_0 \mathbb{p}_1,
\qquad\qquad\mathbb{f}_2=\frac{1}{3} \mathbb{f}_0 \left(\mathbb{p}_1^2-3 \mathbb{p}_2\right),
\qquad\qquad\mathbb{f}_3=\mathbb{f}_3,\nonumber\\
\fl\mathbb{f}_4=\frac{1}{72} \left\{\mathbb{f}_0 \Bigl[\mathbb{p}_1^4-19 \mathbb{p}_1^2\mathbb{p}_2+6 \left(\mathbb{p}_2^2+6 \mathbb{p}_4\right)\Bigr]-18 \mathbb{f}_3 \mathbb{p}_1\right\},...,
\end{eqnarray}
where the first five coefficients of the function $\mathbb{p}(r)$~(\ref{f11}) are given by
\begin{eqnarray}
\fl\mathbb{p}_0=2,\quad~~
\mathbb{p}_{1}=\mathbb{p}_{1},
\quad~~\mathbb{p}_2=\frac{1}{3}\mathbb{p}_1 \left(\mathbb{p}_1-18\right),
\quad~~\mathbb{p}_3=\frac{1}{6} \mathbb{p}_1 \left(\mathbb{p}_1^2+\mathbb{p}_2\right),
\quad~\mathbb{p}_{4}=\mathbb{p}_{4}.
\end{eqnarray}
Above values $\mathbb{f}_{0}$, $\mathbb{f}_{3}$, $\mathbb{p}_{1}$, and $\mathbb{p}_{4}$  are four free parameters and the constant $f_{0}$ in~(\ref{intfr}) becomes
\begin{eqnarray}\label{}
f_{0} =\frac{2520-4609 \mathbb{f}_0\mathbb{p}_1}{2520}.
\end{eqnarray}
However, we find out that, the last equation~(\ref{monopolefield3}) is satisfied if and only if the coefficient
(\ref{fibölüfi}) is equal to $0$. Hence, we get
\begin{eqnarray}\label{}
\mathbb{p}_{1}=\mathbb{p}_{2}=\mathbb{p}_{3}=\mathbb{p}_{4}=...=0,
\end{eqnarray}
which  not only fix the $\mathbb{f}_{1}=\mathbb{f}_{2}=\mathbb{f}_{4}=\mathbb{f}_{5}=...=0$ and the value of constant $f_{0}$ as $f_{0}=1$ but also make the problematic term,  $\mathbb{f}_{1}r^{-1}$, of the $\mathbb{f}(r)$ in ~(\ref{sigmaeksiiki1}), which turns into a logarithmic term,  $\mathbb{f}_{1}\ln r$, in $f(r)$, surprisingly vanish. Finally, the metric function  $f(r)$~(\ref{intfr}) and warping function (\ref{sigmaeksiiki2}) reduce to two simple sets as follows:
\begin{eqnarray}
f(r)&=&1-\frac{\mathbb{f}_{0}}{r}+\frac{\mathbb{f}_{3}}{2}r^{2},\label{second1}\\
\varphi(r)&=&\varphi_{0},\label{fisıfırolunca}
\end{eqnarray}
which means that the warping function is just a constant, $\varphi_{0}$. Assuming $\mathbb{f}_{0}=2M$ and $\mathbb{f}_{3}=-(2/3)\Lambda$, one writes
\begin{eqnarray}\label{bhs}
f(r)=1-\frac{2M}{r}-\frac{1}{3}\Lambda r^{2},
\end{eqnarray}
where the two constants, $M$ and $\Lambda$, may be interpreted as a mass and a cosmological constant of the static black hole, respectively, but this time $\Lambda$ is arbitrary. In the 4D WY limit, several authors~\cite{fairchild76,pavelle1,pavelle3,ni,Hsu} have already  obtained  this solution, which is nothing but the famous neutral Schwarzschild black hole  solution in the de Sitter space-time if $\Lambda >0$. If we assume that $\Lambda < 0$, we obtain the Kottler~\cite{Kottler} solution (also called Schwarzschild-anti-de Sitter), and by setting $\Lambda=0$, we just have the Schwarzschild solution. It is well known that the roots of the black hole solution~(\ref{bhs}) can be classified into three cases:
\begin{eqnarray}
\cases{
\mbox{two real positive zeros},~~ r_{+}<r_{c}, & if~~ $0<9M^{2}\Lambda<1,$\\
\mbox{the extreme case},~~r_{+}=r_{c}=3M, & if~~ $9M^{2}\Lambda=1,$\\
\mbox{no real positive zeros}, & if~~ $9M^{2}\Lambda>1,$\\
}
\end{eqnarray}
where $M>0$ (see, e.g.,~\cite{Molina}). However, Einstein’s vacuum theory ${\hat{\mathcal{R}}}_{A}=\hat\Lambda\hat E_{A}$ gives the 4D solution~(\ref{bhs}) together with  $\hat\Lambda=\Lambda=1/2l^2\varphi_{0}^2$ if we start with assuming the ansatzes (\ref{monopole}) and (\ref{fisıfırolunca}). Hence, the solution~(\ref{bhs}) seems to be more general than Einstein’s vacuum solution. Consequently, the (4+3)D non-Abelian WYKK theory with the Wu$-$Yang ansatz admits the 4D charged and uncharged Schwarzschild-(anti-)de Sitter solutions  (\ref{rnds}) and~(\ref{bhs}), respectively,   if $\varphi(r)=constant$.  Let us also remark that the invariants in this case become
\begin{eqnarray}\label{kis}
\hat{\mathcal{ K}}=\frac{3}{8(l\varphi_{0})^{4}}+\frac{4}{3}\Lambda^{2}+\frac{24M^{2}}{r^{6}},\qquad\qquad
\hat{\mathcal{ R}}=\frac{3}{2(l\varphi_{0})^{2}}+4\Lambda,
\end{eqnarray}
which confirm that
\begin{eqnarray}\label{}
\lim_{r\rightarrow 0}\hat{\mathcal{ K}}\rightarrow\infty,\qquad\qquad\qquad
\lim_{r\rightarrow\infty}\hat{\mathcal{ K}}=\frac{3}{8(l\varphi_{0})^{4}}+\frac{4}{3}\Lambda^{2},
\end{eqnarray}
and the nonzero components of the stress-energy tensor turn out to be
\begin{eqnarray}\label{set00}
\hat T_{tt}=-\hat T_{rr}=-\hat T_{\theta\theta}=-\hat T_{\phi\phi}=\frac{3}{(2l\varphi_{0})^{4}},\nonumber\\
\hat T_{\Theta\Theta}=\hat T_{\Phi\Phi}=\hat T_{\psi\psi}=\frac{1}{(2l\varphi_{0})^{4}}-\frac{2}{3}\Lambda^{2}-\frac{12M^{2}}{r^{6}}.
\end{eqnarray}
Thus, we have a diagonal energy-momentum tensor. At the origin $r=0$, the Kretschmann invariant and the $3$D parts of the stress-energy tensor have a pole, whereas the curvature scaler is finite everywhere. This information suggests that there is a curvature singularity, at $r = 0$, of the solution.

 \subsubsection{The  $\sigma_{2} =-5$ and $\mathbb{p}_0=5$ case}

Let us search for black hole solutions by assuming that  $\sigma_{2} =-5$ and $\mathbb{p}_0=5$. Thus, from the equations~(\ref{fere}) and~(\ref{scalarserie}) we have
\begin{eqnarray}\label{sigmaeksibes1}
\fl \mathbb{f}(r)
=\frac{\mathbb{f}_{0}}{r^{5}}+\frac{\mathbb{f}_{1}}{r^{4}}
+\frac{\mathbb{f}_{2}}{r^{3}}+\frac{\mathbb{f}_{3}}{r^{2}}
+\frac{\mathbb{f}_{4}}{r}+\mathbb{f}_{5}+\mathbb{f}_{6}r+...,\\
\fl\varphi(r)=\varphi_{0}\left[r+\frac{\mathbb{p}_{1}}{3}r^{2}
+\frac{(\mathbb{p}^{2}_{1}+3\mathbb{p}_{2})}{18}r^{3}
+\frac{(\mathbb{p}^{3}_{1}+9\mathbb{p}_{1}\mathbb{p}_{2}+18\mathbb{p}_{3})}{162}r^{4}+...\right].\label{sigmaeksibes2}
\end{eqnarray}
It is straightforward to verify again that the first field equation~(\ref{monopolefield1}) restricts the coefficients of the  $\mathbb{f}(r)$  in the expansion~(\ref{sigmaeksibes1}) as follows:
\begin{eqnarray}\label{parameters}
\fl\mathbb{f}_{0}=\mathbb{f}_{0},\qquad\qquad
\mathbb{f}_1=-\frac{7}{5} \mathbb{f}_0 \mathbb{p}_1,\qquad\qquad
\mathbb{f}_2=\frac{1}{120} \mathbb{f}_0 \left(121 \mathbb{p}_1^2-105 \mathbb{p}_2\right),\nonumber\\
\fl\mathbb{f}_3=-\frac{1}{1080}\mathbb{f}_0 \left(549 \mathbb{p}_1^3-1453 \mathbb{p}_1\mathbb{p}_2 +840 \mathbb{p}_3\right),\nonumber\\
\fl\mathbb{f}_4=\frac{1}{8640}\mathbb{f}_0 \left(1833 \mathbb{p}_1^4-9934 \mathbb{p}_1^2\mathbb{p}_2 +11712 \mathbb{p}_1\mathbb{p}_3 +4365\mathbb{p}_2^2-7560 \mathbb{p}_4\right),\nonumber\\
\fl\mathbb{f}_5=\frac{1}{129600}\mathbb{f}_0 \Biggl[-12105 \mathbb{p}_1^5+113062  \mathbb{p}_1^3\mathbb{p}_2-205176 \mathbb{p}_1^2\mathbb{p}_3+\nonumber\\
+3\mathbb{p}_1 \left(89496 \mathbb{p}_4-50519 \mathbb{p}_2^2\right) +360 \left(491 \mathbb{p}_2 \mathbb{p}_3-504 \mathbb{p}_5\right)\Biggr],\nonumber\\
\fl\mathbb{f}_{6}=\mathbb{f}_{6},
\end{eqnarray}
where $\mathbb{f}_{0}$ and $\mathbb{f}_{6}$  are arbitrary constants
and the integration constant, $f_{0}$, in~(\ref{intfr}) turns into following explicit expression:
\begin{eqnarray}\label{fsıfır}
\fl f_{0}=\frac{1}{21772800}\Biggl[5443200+\mathbb{f}_0(8219607 \mathbb{p}_1^4-53024506 \mathbb{p}_1^2 \mathbb{p}_2 +72306048 \mathbb{p}_1 \mathbb{p}_3+\nonumber\\
+27177435 \mathbb{p}_2^2-53214840 \mathbb{p}_4)\Biggr].
\end{eqnarray}
Conversely, the second field equation~(\ref{monopolefield2}) is satisfied if
$\mathbb{p}_{1}=\mathbb{p}_{2}=\mathbb{p}_{3}=\mathbb{p}_{4}=\mathbb{p}_{5}=...=0$,
which means that the recursion equation implies
$\mathbb{f}_{1}=\mathbb{f}_{2}=\mathbb{f}_{3}=\mathbb{f}_{4}=\mathbb{f}_{5}=\mathbb{f}_{7}=...=0$, in the equation~(\ref{parameters})
with $f_{0} =1/4$ in the equation~(\ref{fsıfır}). In this fashion, the problematic term,  $\mathbb{f}_{4}r^{-1}$, of the $\mathbb{f}(r)$ in ~(\ref{sigmaeksibes1}) suddenly disappears again. We also obtain the condition $\varphi_{0}=\pm1/\sqrt{2}l$ from the last field equation~(\ref{monopolefield3}), where $\varphi'(r)/\varphi(r)\neq 0$. Finally, by substituting these results into two unknown functions, $f=f(r)$~(\ref{intfr}) and $\varphi=\varphi(r)$~(\ref{sigmaeksibes2}), gives
\begin{eqnarray}\label{}
f(r)=\frac{1}{4}-\frac{\mathbb{f}_{0}}{4r^{4}}+\frac{\mathbb{f}_{6}}{2}r^{2},\label{second2}\\
\varphi(r)=\pm\frac{1}{\sqrt{2}l}r,\label{nconst}
\end{eqnarray}
where the warping function is free from singularities and has two branches with a ``$+$'' or ``$-$'' sign. It is not satisfying  the boundary condition $\lim_{r\rightarrow\infty}\varphi(r)=0$, as well. We can say that the warping function~(\ref{nconst}) eliminates the constant radius term, $l$, in the metric~(\ref{metricc2}) and defines a variable radius for three-sphere where the external and internal spaces are coupled. Only when $\varphi(r)=\varphi_{0}$,  these two spaces are decoupled and the 4D solutions~(\ref{rnds}) and~(\ref{bhs}) appear from the higher-dimensional theory.

To interpret the physical properties of the metric function
$f(r)$ in the equation~(\ref{second2}), we can provide a brief review on the D-dimensional (anti-)de Sitter black holes solutions in the Einstein$-$Maxwell gravity, where the metric takes the ansatz
\begin{eqnarray}\label{newmet}
ds^{2}=-f(r)dt^{2}+\frac{dr^{2}}{f(r)}+\varphi(r)^{2}d\Omega^{2}_{k},
\end{eqnarray}
with the line element of a $(D-2)$-dimensional hypersurface, $d\Omega^{2}_{k}$. The scalar curvature is equal to $(D-2)(D-3)k$, where the value of parameter $k=1$, $-1$, or $0$. For choosing warping function $\varphi(r)=r$, the charged black hole solution is given by (see, e.g.,~\cite{Liu2004,Astefanesei2004,Brihaye2009})
\begin{eqnarray}\label{higherd}
f(r)=k-\frac{2M}{r^{D-3}}+\frac{Z^{2}}{r^{2(D-3)}}-\frac{2}{(D-1)(D-2)}\Lambda r^{2},
\end{eqnarray}
where, as usual, $M$ and $Z$ are the mass and the charge parameters and the $\Lambda$ can be interpreted as a cosmological constant, respectively. Furthermore, for no warping $\varphi(r)=1$ and $Z=0$,  a class of solution is given by~\cite{Cadeau2000}
 \begin{eqnarray}\label{class2}
f(r)=C-\frac{2}{3}\Lambda r^{2},
\end{eqnarray}
where $C$ is a constant of integration. As can be easily seen,  solution~(\ref{bhs}) is different from the above solution with the nonzero mass term  and  $\varphi(r)=\varphi_{0}$, where $\varphi_{0}$ is any constant. That is more general than value $1$.

Thus, if we define $\mathbb{f}_{0}=8M$ and $\mathbb{f}_{6}=-(2/15)\Lambda$ in the equation~(\ref{second2}) inspired by the equation~(\ref{higherd}) for the $Z=0$ and $D=7$, the new exact solution can be written as follows:
\begin{eqnarray}\label{bhs1}
f(r)=\frac{1}{4}-\frac{2M}{r^{4}}-\frac{1}{15}\Lambda r^{2}.
\end{eqnarray}
The equation is formally equivalent to the purely uncharged (anti-)de Sitter metric, and it seems to be a special case of the topological seven-dimensional Schwarzschild$-$de Sitter solution  with $\Lambda>0$ and $k=1/4$.
Rewriting $f(r)=0$ by~(\ref{bhs1}), we have
\begin{eqnarray}\label{roots}
\Delta(r)=\frac{1}{15}\Lambda r^{6}-\frac{1}{4}r^{4}+2M=0,
\end{eqnarray}
which has six horizons, $r_{i}$, at most including the complex or negative roots correspond to the virtual (unphysical) horizons. By applying Vieta’s theorem on the equation~(\ref{roots}), we find that
\begin{eqnarray}\label{Vieta}
\sum_{1\leq i<j\leq6}r_{i}r_{j}=-\frac{15}{4\Lambda},\qquad\qquad\qquad\qquad
\prod_{i=1}^{6}r_{i}=30\frac{M}{\Lambda}.
\end{eqnarray}
For instance, the positive event horizon can be written as follows:
\begin{eqnarray}\label{}
r_{+}=\sqrt{\frac{1}{12 \Lambda }(\Omega+15+\frac{225}{\Omega})},
\end{eqnarray}
where
\begin{eqnarray}\label{}
\Omega=3 \sqrt[3]{125-960 \Lambda ^2 M+40\sqrt{6\Lambda ^2 M \left(96 \Lambda ^2 M-25\right)}}.
\end{eqnarray}
To avoid the space-like naked singularity, we must again be careful when choosing the constants $M$ and $\Lambda$. The Kretschmann and Ricci scalars are calculated using the equation~(\ref{bhs1}) as follows:
\begin{eqnarray}\label{kas}
\hat{\mathcal{K}}=\frac{14}{75}\Lambda^{2}+\frac{9}{4r^{4}}
+\frac{300M^{2}}{r^{12}},\qquad\qquad\qquad
\hat{\mathcal{R}}=\frac{14}{5}\Lambda.
\end{eqnarray}
It is essential here to note that, the nonzero curvature scalar can be written as
\begin{eqnarray}
\hat{\mathcal{R}}=\frac{2D\Lambda}{(D-2)},
\end{eqnarray}
for the metric~(\ref{newmet}) in the higher-dimensional theory (see, e.g.,~\cite{Hendi2011,Sheykhi2012}). The $\hat{\mathcal{R}}$ in the equation~(\ref{kas}) is compatible with the above result for the $D=7$ in this fashion. The covariantly conserved and diagonal energy-momentum tensor also reduces to the following components:
\begin{eqnarray}\label{set01}
\hat T_{tt}=-\hat T_{rr}=\frac{1}{25}\Lambda^{2}+ \frac{9}{8r^{4}}-\frac{90M^{2}}{r^{12}},\nonumber\\
\hat T_{\theta\theta}=\hat T_{\phi\phi}=-\frac{1}{25}\Lambda^{2}+ \frac{3}{8r^{4}}-\frac{126M^{2}}{r^{12}},\nonumber\\
\hat T_{\Theta\Theta}=\hat T_{\Phi\Phi}=\hat T_{\psi\psi}=-\frac{1}{25}\Lambda^{2}- \frac{5}{8r^{4}}-\frac{126M^{2}}{r^{12}}.
\end{eqnarray}
The curvature invariant and the energy-momentum tensors, again, are all finite everywhere except at $r=0$. At $r\rightarrow 0$, $\hat{\mathcal{ K}}$ goes to $\infty$, but at $r\rightarrow\infty$, $\hat{\mathcal{ K}}$ goes to $\hat{\mathcal{ K}}=(14/75)\Lambda^{2}$ and $\varphi(r)$ diverges, which confirms that there is a curvature singularity at $r=0$.

Both for $\sigma_{1} =1$, $\mathbb{p}_0=2$ and $\sigma_{1} =1$, $\mathbb{p}_0=5$ cases, conversely, we find that the first solution is of the form, $f_{1}(r)=f_{0}$. We can see that the first solution, $f_{1}(r)$, appears in the second solutions~(\ref{second1}) and (\ref{second2}), respectively. This means that if we start with a smaller root of the indicial equation, we can obtain~(\ref{second1}) and (\ref{second2}) as the general solution despite not including a logarithmic term.

Let us finally remark that it is quite natural to consider the case where the time components of the gauge fields, as well as the warping function, survive and the space components of the gauge fields vanish. Because of  the coupled field equations, there are no solutions in this case (see, e.g., ~\cite{Winstanley121}).

 \section{Conclusions}

In this paper, we have combined the WY gravitational gauge theory with the KK theories of unified gravity against the background of the non-Abelian gauge fields. This is, from a theoretical point of view, a new and rather appealing approach in the literature. The theoretical description  was reduced from seven to four dimensions by considering  the exterior differential forms that provide the more transparent equations along with shortcut calculations. The corresponding equations of motion (\ref{a4})--(\ref{rfe6}) were not only mathematically more complicated than those of the non-Abelian version of the KK theory that comes from Einstein's gravity, i.e., the $\{\mathcal{P}_{\mu\nu}, \mathcal{Q}_{i\nu}, \mathcal{U}_{ij}\}$-set in (\ref{Einsteinian}) but also included  additional  physical  information because of the more general field equations. We could easily obtain, for instance, the YM force density-like terms in equation~(\ref{lfd}), which only appears in geodesic equations of the ordinary KK theories, and the usual $4$D continuity equation without making any further assumptions. Furthermore, the right side of the field equation (\ref{a4}), i.e., is the $S_{\mu\nu\sigma}$ term~(\ref{rfe16}), may conveniently be interpreted as the source term of the $4$D WY gauge theory, that is missing in the literature. The reduction mechanism transformed the higher-dimensional sourceless WY theory  into the $4$D WY theory with a source term  in the usual $4$D space-time. In fact, the same situation has already been discussed for the $5$D version of the presented theory with the $U(1)$ Abelian case~\cite{halil13}. A similar accomplishment has already been obtained by ordinary KK theories, in which the $5$D vacuum equations were reduced to the $4$D Einstein$-$Maxwell-scalar system. This is sometimes called the KK miracle.

We have also constructed the exact static (anti-)dyon~(\ref{rnds}) and  pure magnetic (anti-)monopole solutions~(\ref{bhs}) and~(\ref{bhs1}) that appear in the literature for the first time in connection with the WYKK theory with the generalized Wu$-$Yang ansatz and some different types of warping function, keeping in mind the spherically symmetric  stationary  external and internal space-times with the product topology  $M_{4}\times S^{3}$. For the sake of simplicity, we concentrated only on the pure internal symmetry group ${\mathcal{G}}=SU(2)$, which led to the mixing of the space-time and color indices. These strong restrictions and highly symmetric assumptions were necessary because of  the complex nature of the considered model and the fact that it is already enormously difficult to find a solution in higher dimensions because there is no general method for the same.

To analytically discover regular solutions, two special cases with $J(r)\neq0$, $K(r)=K$ and $J(r)=0$, $K(r)=K$ were also considered as a type of strategy. In this manner, we obtained the singular Wu$-$Yang-type (anti-)dyon solution that corresponded to the RNdS-like black hole solution~(\ref{rnds}), which is in a more desirable and suitable form than  both Perry's (\ref{rnds1}) and Angus' (\ref{rnds2}) solutions for the first case. Additionally, its spherically symmetric gauge potentials (\ref{pote}) are more general and non-Abelian than both solutions (\ref{gfperry}) and (\ref{gfangus}) and carry unit electric and magnetic charges (\ref{dyoncharge}) in the 4D. The overall results are summarized in Table $1$. It was demonstrated in this approach that the cosmological constant naturally appeared as an integration constant (i.e., ``it is a free parameter unrelated to the vacuum energy of quantum fields'' according to Cook~\cite{cook}) and that it was associated with the constant warping function as well as with the three-sphere radius. The general structure of this regular black hole was also studied in detail using its mathematical similarity to the well-known RNdS black hole solution.

Whether the solution is a monopole or an antimonopole may depend on the sign of the massive scalar Higgs field~\cite{Marciano,Marciano1}, on the sign of the Bogomol'nyi equations~\cite{Rossi}, or on the $\phi$-winding number, which can be a positive or negative integer of the internal  space~\cite{Teh}. We have shown in this study that the sign of the internal space-squashing parameter $\ell$ determines  whether the solution is a dyon/monopolo or an antidyon/antimonopole in  (\ref{dyoncharge}) and (\ref{dyoncharge2}).

For the Brans$-$Dicke-type case, to solve the coupled linear differential equations~(\ref{monopolefield1})--(\ref{monopolefield3}), we used the Frobenius method, which  not only gives all possible solutions but also is the most suitable approach to solve the first field equation~(\ref{monopolefield1}), mathematically. In this  regard, the nontrivial solutions are the 4D (\ref{bhs}) and 7D (\ref{bhs1}) neutral Schwarzschild-like black hole solutions in the (anti-)de Sitter space-time, respectively. The external and internal spaces are coupled if $\varphi(r)=\pm(1/\sqrt{2}l)r$ and the 7D non-Abelian WYKK theory gives 7D solutions. When $\varphi(r)=\varphi_{0}$, these two spaces are decoupled, giving 4D solutions that are not only different from the (\ref{class2}) but also more general than Einstein’s vacuum solution.  Therefore, the existence of nonconstant warping function in  (\ref{nconst}) is essential for finding new-type black hole solutions (\ref{bhs1}) (see also ~\cite{kuyrukcu}). Hence, the warping function plays very important roles in the WY theory just as
the scalar fields do in the supergravity theories.

Conversely, the corresponding gravitational stress-energy tensors of these solutions were also analyzed. Especially,~(\ref{set00}) and~(\ref{set01}) are just diagonal and formally similar to that of Maxwell's theory of electromagnetism. The $\hat T_{AB}$ in equations (\ref{angle3})--(\ref{angle1}) can also be diagonalized, if we use proper values for the $r$, $\theta$, and $\phi$. Also, we determined  Kretschmann scalars to present that there is a curvature singularity at $r=0$. In other words, these solutions are interpreted as  charged or uncharged black hole solutions with the cosmological constant.

It would be interesting to discover the Abelian solutions of the $5$D version of this alternative gravitational model in  both the ordinary $4$D and $5$D space-times. We may consider the $U(1)$ Abelian ansatz from the non-Abelian gauge by assuming that only the ${{A}}^{\alpha}_{a}(\vec{x})$ component of the gauge field survives, ${{A}}^{\alpha}_{a}(\vec{x})=\delta^{\alpha}_{3}{{A}}_{a}^{Dirac}(r,\theta)$. Obtaining the $4$D field equations from the Weyl-rescaled reduced action and investigating whether this nonlinear coupled system of partial differential equations possesses a well-posed initial value formulation as other higher-derivative gravity theories~\cite{Noakes} are other important mathematical challenges. A detailed discussion of these problems will be presented in a forthcoming study.

\ack

The author would like to thank the anonymous referees for their helpful and thorough comments on an earlier version of the manuscript.

\end{document}